%% file: main.tex
\documentclass[11pt]{article}

\usepackage{graphicx}
\usepackage[a4paper, total={7in, 10in}]{geometry}
\usepackage{lmodern}

\usepackage{chemformula} 
\usepackage[T1]{fontenc} 
\usepackage{siunitx} 
\usepackage{comment} 
\usepackage{multirow}
\usepackage{booktabs} 
\usepackage[version=4]{mhchem}
\usepackage{gensymb}
\usepackage{soul}
\usepackage{amsmath, amsfonts, amssymb}
\usepackage{titlesec}
\usepackage{authblk}

\titleformat{\section}
  {\normalfont\Large\bfseries}{\thesection}{1em}{}

\usepackage{array}                    
\newcolumntype{C}[1]{>{\centering\arraybackslash}p{#1}}  
\newcolumntype{R}[1]{>{\raggedleft\arraybackslash}p{#1}} 

\include{defs}

\title{ \rule{\textwidth}{2pt} 
{\bf Liquid phase fast electron tomography unravels the true 3D structure of colloidal assemblies} \\[-1.5ex]
\rule{\textwidth}{2pt} 
}

\author[$1,\dagger$]{Daniel Arenas Esteban}
\author[$1,2,\dagger$]{Da Wang}
\author[$1,3,\dagger$]{Ajinkya Kadu}
\author[$1$]{Noa Olluyn}
\author[$4,5,6$]{Ana S{\'a}nchez Iglesias}
\author[$7$]{Alejandro Gomez-Perez}
\author[$8$]{Jes{\'u}s Gonzalez Casablanca}
\author[$7$]{Stavros Nicolopoulos}
\author[$4,5,9,10,*$]{Luis M. Liz-Marz{\'a}n}
\author[$1,*$]{Sara Bals}

\affil[1]{Electron Microscopy for Materials Science (EMAT) and NANOlab Center of Excellence, University of Antwerp, Groenenborgerlaan 171, 2020 Antwerp, Belgium.}
\affil[2]{Guangdong Provincial Key Laboratory of Optical Information Materials and Technology, Institute of Electronic Paper Displays, South China Academy of Advanced Optoelectronics, South China Normal University, Guangzhou 510006, China.}
\affil[3]{Centrum Wiskunde \& Informatica (CWI), Amsterdam, The Netherlands.}
\affil[4]{CIC biomaGUNE, Paseo de Miramon 182, 20009 Donostia-San Sebasti\'an, Spain.}
\affil[5]{Centro de Investigaci\'on Biom\'edica en Red, Bioingenier\'ia, Biomateriales y Nanomedicina, CIBER-BBN, Paseo de Miramon 182, 20009 Donostia-San Sebasti\'an, Spain.}
\affil[6]{Centro de F\'isica de Materiales CSIC-UPV/EHU Paseo Manuel de Lardizabal 5, 20018 Donostia-San Sebasti\'an, Spain.}
\affil[7]{NanoMegas SRPL, Rue \`Emile Claus 49 bte 9, Brussels, 1050, Belgium.}
\affil[8]{Universidad Rey Juan Carlos, Centro de Apoyo Tecnol\'ogico, c/Tulipan s/n 28933 Madrid, Spain.}
\affil[9]{Ikerbasque, Basque Foundation for Science, 48013 Bilbao, Spain.}
\affil[10]{Cinbio, Universidade de Vigo, 36310 Vigo, Spain.}
\affil[$\dagger$]{These authors contributed equally to this work}

\date{}

\begin{document}

\maketitle


\begin{abstract}
Electron tomography has become a commonly used tool to investigate the three-dimensional (3D) structure of nanomaterials, including colloidal nanoparticle assemblies. However, electron microscopy is typically carried out under high vacuum conditions. Therefore, pre-treatment sample preparation is needed for assemblies obtained by (wet) colloid chemistry methods, including solvent evaporation and deposition on a solid TEM support. As a result of this procedure, changes are consistently imposed on the actual nanoparticle organization. Therefore, we propose herein the application of electron tomography of nanoparticle assemblies while in their original colloidal liquid environment. To address the challenges related to electron tomography in liquid, we devised a method that combines fast data acquisition in a commercial liquid-TEM cell, with a dedicated alignment and reconstruction workflow. We present the application of this method to two different systems, which exemplify the difference between conventional and liquid tomography, depending on the nature of the protecting ligands. 3D reconstructions of assemblies comprising polystyrene-capped Au nanoparticles encapsulated in polymeric shells revealed less compact and more distorted configurations for experiments performed in a liquid medium compared to their dried counterparts. On the other hand, quantitative analysis of the surface-to-surface distance of self-assembled Au nanorods in water agrees with previously reported dimensions of the ligand layers surrounding the nanorods, which are in much closer contact when in similar but dried assemblies. This study, therefore, emphasizes the importance of developing high-resolution characterization tools that preserve the native environment of colloidal nanostructures.
\end{abstract}


\section{Introduction} 
\label{sec:Intro}  

The most characteristic feature of nanomaterials is the stark dependence of their properties on the size and shape of the nanostructured material. However, manipulation of the properties of nanomaterials can also be achieved by tuning interparticle distance and relative orientation\cite{guerrero2012molecular}. In this context, a wide variety of techniques have been devised toward obtaining nanostructured materials with well-defined dimensions and interparticle arrangements. Although top-down methods, such as e-beam lithography, can be used to design nanostructures with high precision, these are typically limited to two dimensions (2D). On the other hand, bottom-up strategies based on colloid science can be employed to obtain 2D or 3D assemblies comprising combinations of (equal or dissimilar) nanosized particles, with distinct properties determined by the size, shape, and arrangement of the constituting elements\cite{GlotzerSolomon2007, boles2016self}. The formation and behavior of such assemblies are governed by interaction forces between nanoparticles (NPs), typically mediated by surface charge, ligands, and the solvent. As such, colloids have long been used as model systems for investigating fundamental phenomena in soft condensed matter, such as nucleation and phase transitions. In the context of nanoscale materials, assemblies additionally play a key role in shaping functional (meta)materials at various scales.

Therefore, understanding the formation mechanisms and structure-determined properties of colloidal assemblies requires quantitative 3D structural characterization, including measurements of interparticle distances and packing\cite{wang2021binary,wang2018interplay,wang2022structural}. Although bulk scattering methods have often been employed with great success and high precision, they can only provide average information over a huge number of individual particles or clusters thereof. Accurate and comprehensive information can be obtained by studying individual nanostructures in 3D, for which electron tomography (ET) in scanning transmission electron microscopy (STEM) has become an essential tool\cite{miao2016atomic,midgley2009electron,altantzis2021optimized,bals2014three}. Apart from providing detailed reconstruction images (or movies), recent improvements have been made to ET, both related to the acquisition and to the reconstruction process, which, e.g., enable us to extract the positions of individual nanoparticles, even in very large and/or dense nanoassemblies, where missing wedge and streaking artifacts are likely to hinder relevant features\cite{zanaga2016quantitative,kavak2023quantitative}.

However, all these investigations have been performed under conventional conditions in a TEM, including ultra-high vacuum. Samples for ET are therefore typically prepared by dropcasting the colloidal dispersion on a TEM grid. A problem that has often been overlooked during this process is related to the presence of soft materials within colloidal assemblies, such as ligands and polymers\cite{lyu2023electron}. As a result, the drying process may result in deformations of the assemblies, either by evaporation of the remaining solvent or by contact with the support (grid), thereby altering their original 3D configuration\cite{marchetti2023templated}. To mitigate this effect, cryogenic electron tomography (cryo-TEM) and freeze-drying have been used, but the experimental environment can still lead to subtle changes in the 3D structures\cite{kumar2018detection,de2015entropy,zanaga2016quantitative,wang2021quantitative}. It is therefore important to develop 3D characterization approaches based on ET that enable the investigation of colloidal assemblies in their natural environments, such as water or a different solvent.

Recent advancements in liquid cell electron microscopy have yielded new insights into nanomaterial dynamics and structure in liquid environments\cite{de2019resolution,ross2015opportunities,de2016investigating}. Initial attempts utilized amorphous silicon nitride ($Si_xN_y$) microfluidic chambers as liquid cells (LCs), but the holders based on such chambers often have a limited tilt range (restricted to $\pm 30^\circ$). For ET, where a sufficient angular sampling is desired, such a limited tilt range can cause missing wedge artifacts, compromising reconstruction accuracy. Additionally, the presence of the liquid layer and relatively thick $Si_xN_y$ windows frequently reduces the signal-to-noise ratio (SNR) in the tilt-series projection images, especially at higher angles where the total effective thickness increases. To address these limitations, graphene liquid cells (GLCs) have been employed to enhance SNR while allowing the study of nanomaterial growth, self-assembly, and dynamics\cite{park2021graphene}. Using the 3D SINGLE methodology, Park \textit{et al.} demonstrated that the 3D structures of NPs in GLC can be characterized by observing their translational and rotational motions in a liquid environment\cite{park20153d}. Using advanced reconstruction algorithms, the structural disparities between individual NPs could be discerned, even at the atomic scale\cite{kim2020critical}. However, the spatial constraints of the GLC  (commonly up to 100 nm) make the single-particle method sub-optimal for 3D characterization of significantly larger colloidal assemblies\cite{yang2019dynamic,keskin2021verification}.

We present herein an advanced Liquid-Phase (LP) fast electron tomography workflow to characterize 3D structures of colloidal assemblies in their native environments. This approach applies \textit{fast electron tomography}\cite{albrecht2020fast,vanrompay2021fast,koneti2019fast}, a recently proposed technique to significantly reduce the acquisition time for ET tilt series, using a commercially available LC chip. To overcome experimental challenges such as limited tilt range, image distortion, environmental background noise, and potential intra-LC sample movement, advanced image processing techniques and a dedicated reconstruction algorithm are proposed\cite{batenburg2011dart,donoho2006compressed}, as well as the use of a novel prototype LC-chip (Tomochip) that allows a higher tilt angle. To illustrate the importance of 3D characterization by LP fast electron tomography, we investigated the structure of colloidal clusters comprising hydrophobic Au NPs surrounded by a block-copolymer shell that provides colloidal stability in water. Our analyses reveal subtle structural differences when comparing colloidal clusters studied in the liquid phase and in vacuum (\ie, dried state). As a further demonstration of the importance of characterizing nanoassemblies in their native environment, we studied bilayer assemblies of cetyltrimethylammonium bromide (CTAB)-stabilized Au nanorods (NRs) in water, revealing the surface-to-surface distance between Au NRs in water, in agreement with literature values\cite{gomez2012surfactant}. These in situ 3D measurements are in contrast to observations of NR assemblies characterized in a dried state and for which significantly smaller distances were determined. Our results therefore illustrate the importance of performing 3D characterization of NP assemblies in a liquid environment. Based on such advancements, a more comprehensive and accurate 3D analysis of colloidal assemblies in their native conditions becomes possible.


\section{Results} 
\label{sec:Results}

\subsection{Challenges in 3D structural characterization of colloidal clusters by electron tomography}
\label{sec:results:Challenges}

As a model system for colloidal assemblies, we followed our previous work\cite{sanchez2012hydrophobic}, in which the self-assembly of polystyrene (PS)-capped Au NPs was induced by adding water to a dispersion of the (hydrophobic) Au-PS NPs in tetrahydrofuran (THF), and subsequently stabilizing the obtained NP clusters by further addition of a polystyrene-b-polyacrylic acid (PSS-PAA) block copolymer. Whereas the PS block interdigitates with the PS ligands on Au NPs, the PAA block allows redispersion of the protected (hydrophilic) assemblies in water. It should be noted that, aiming to enhance the interdigitation of PS chains between the NPs inside the cluster, slight heating was applied to help expel the remaining THF. Therefore, the NP clusters redispersed in water are expected to be compact and allow minimum internal dynamics. However, this hypothesis could not be tested by standard ET in vacuum because sample preparation would lead to the complete evaporation of any remaining solvent. This effect is likely to increase PS chain interdigitation, thereby further reducing interparticle distance. As a result, our reported tomography reconstructions typically showed a highly regular organization of Au NPs, with interparticle distances regulated by the dimension (molecular weight) of the PS ligands\cite{galvan2014self}.

\begin{figure}[!htb]
    \centering
    \includegraphics[width=0.9\textwidth]{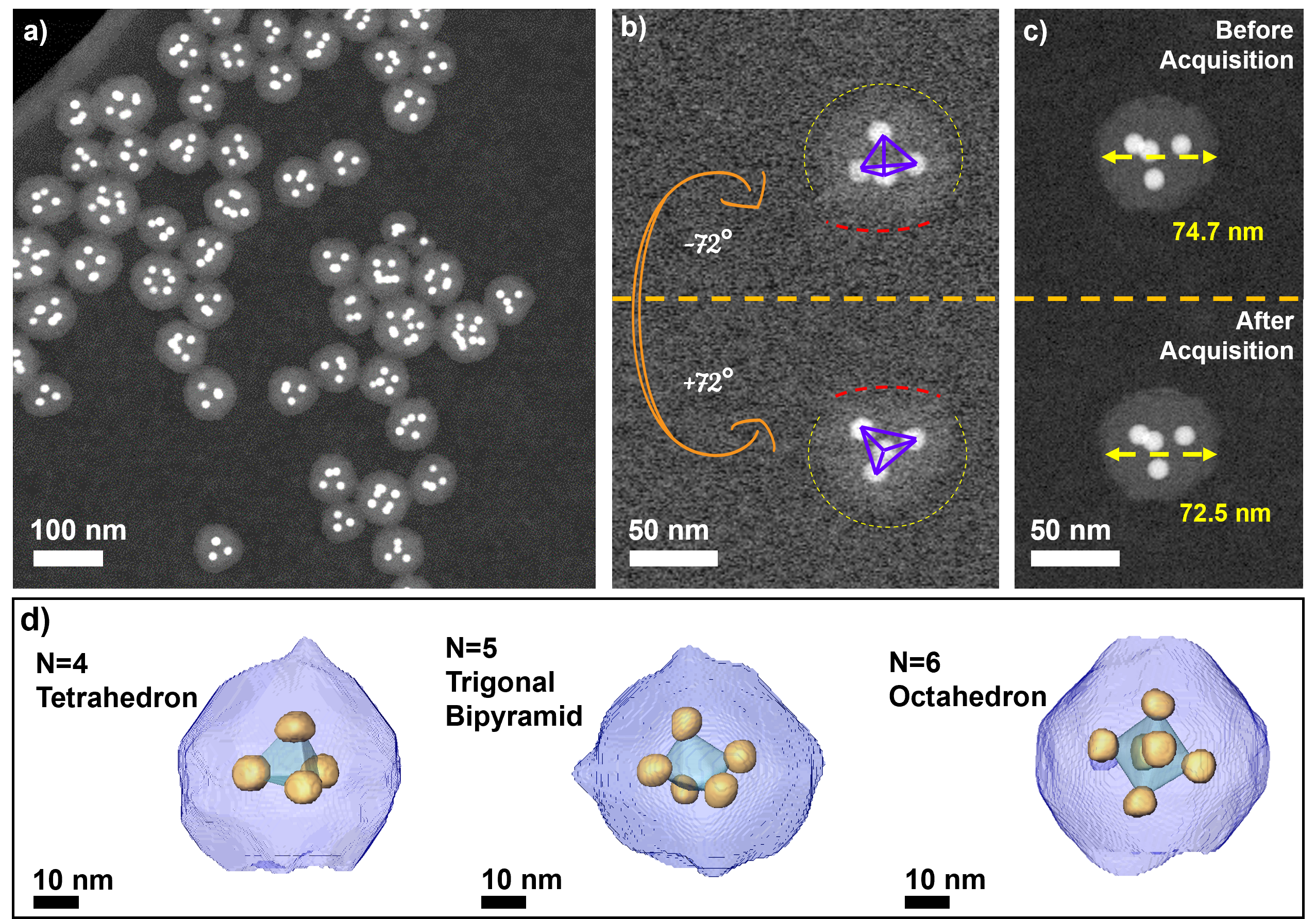}
    \caption{\textbf{Challenges in 3D characterization of colloidal clusters by electron tomography}. 
    a) HAADF-STEM image showing an overview of Au@PS colloidal clusters in vacuum, where the polymer shell can be observed as a grey shadow around the bright NPs.
    b) Observation of the flattening effect of a colloidal cluster at a high-tilting angle.
    c) 2D HAADF-STEM images of a colloidal cluster before (top) and after (bottom) conventional electron tomography tilt series acquisition, indicating volume change. See also Supplementary Movie 1.
    d) 3D reconstructions of colloidal clusters containing 4, 5, and 6 Au NPs \textit{via} fast electron tomography in vacuum. The stacking of Au NPs within the polymeric shells resembles a tetrahedron, a trigonal bipyramid, and an octahedron, respectively. See also Supplementary Movie 2.}
     \label{fig.1} 
\end{figure}

For high-angle annular dark-field (HAADF)-STEM tomography experiments, we prepared colloidal clusters made of 12 nm Au NPs, with an overall average cluster diameter of 80 nm (Fig.~\ref{fig.1}a; synthesis details are provided in the Methods section). We first applied conventional ET in vacuum to Au@PS clusters. For conventional ET, selected clusters comprising 4, 5, or 6 NPs, and encapsulated within polymer shells, were thoroughly dried on a TEM grid and imaged in vacuum. We noted that the colloidal clusters settled onto the TEM grid upon drying, which resulted in a slightly deformed or flattened structure, evident from 2D TEM projections at high tilt angles (Fig.~\ref{fig.1}b). The 2D projections from the tilt series suggest Au NP stacking into polyhedral structures, \eg, tetrahedra for clusters with 4 NPs (Fig.~\ref{fig.1}b and Supplementary Movie 1). Our analysis also showed a 2-3 nm reduction in the overall size of the colloidal clusters post conventional ET experiments (Fig.~\ref{fig.1}c; Supplementary Fig.~\ref{SI_fig_1}a-c), suggesting that the electron beam has a significant impact on the structure, in turn posing additional challenges when attempting to obtain accurate 3D reconstructions using conventional methods (Supplementary Notes).

To mitigate electron beam damage during tilt series acquisition, we employed the fast electron tomography method, maintaining a relatively low electron dose per frame ($\sim 0.46~\text{e}^-/\text{Å}^2$; see Methods section for more details). In this approach, focusing and tracking are executed concurrently while the sample is continuously tilted\cite{albrecht2020fast,koneti2019fast}. We devised advanced image processing and alignment techniques, coupled with a reconstruction algorithm, to address distortions from continuous tilting and the challenges of a low electron dose. Further details are provided in the Methods section. The 3D reconstructions of colloidal clusters comprising 4, 5, and 6 Au NPs distinctly showcased tetrahedral, trigonal bipyramidal, and octahedral arrangements (Fig.~\ref{fig.1}d and Supplementary Movie 2). Post fast electron tomography, although the overall size of the colloidal clusters remained consistent, flattening of the polymeric shell was observed (Fig.~\ref{fig.1}d and Supplementary Movie 2). It is therefore likely that the capillary forces exerted during sample drying affected the polymeric shell structures, again posing challenges for precise 3D reconstruction. Consequently, these findings underscore the need to investigate the 3D arrangement of Au NPs in colloidal clusters in their native environment.

Drawing inspiration from the 3D SINGLE methodology developed by Park \textit{et al.}\cite{park20153d,kim2020critical}. we aimed to achieve 3D reconstruction by tracking the Brownian motion of NP clusters in a liquid phase. Given the spatial constraints within the GLC, we housed the colloidal cluster dispersion in the $Si_xN_y$ chamber of a commercial liquid TEM holder (Supplementary Notes). After initiating flow within the commercial LC TEM holder, we attempted to capture the dynamics of the colloidal clusters (Supplementary Fig.~\ref{SI_fig_2}a). Contrary to the expected translational motion due to liquid flow, the colloidal clusters exhibited only minimal degree of rotation within the LC chamber, irrespective of the flow rate. As a result, the acquired angular samplings were inadequate for tomographic reconstruction. Additionally, we noted partial aggregation, which may be attributed to degradation of the protective polymer shell under electron beam exposure (Supplementary Fig.~\ref{SI_fig_2}b-d; Supplementary Movie 3). In summary, the reliability of tomographic reconstructions for colloidal clusters in a liquid setting is compromised by challenges related to limited angular projections and electron beam-induced damage. We therefore developed an optimized workflow for the acquisition, alignment, and reconstruction of accurate 3D representations of colloidal NP clusters, which we present in the following sections.

\subsection{Fast tilt series acquisition for LP fast electron tomography}
\label{sec:results:LP-FET}

To increase the angular sampling during the collection of 2D projection images and thereby minimize electron beam-induced structural damage in a liquid environment, we performed fast electron tomography in a LC. We employed a commercial monolithic LC (K-Kit from Bio MA-TEK) container with a window gap of $0.5~\mu m$ (Fig.~\ref{fig_2}a). This device enables the structural characterization of samples within a liquid environment and can be mounted onto a standard single-tilt holder (Fig.~\ref{fig_2}b, c). In comparison to commercially available LCs with a limited inclination angular range (usually not more than $\pm 30\degree$), our setup reaches a slightly extended total tilt range of approximately $\ 90 \degree$ (\ie, $\pm 45\degree$ from the central axis). To minimize beam damage during the tilt series acquisition, we adopted the fast electron tomography acquisition methodology that we previously used in vacuum conditions, as described above\cite{albrecht2020fast}. The electron doses per frame were set as $0.46~\text{e}^-/\text{Å}^2$ and $2.31~\text{e}^-/\text{Å}^2$ for liquid and vacuum conditions, respectively (Fig.~\ref{fig_2}d). Due to the reduced beam current employed to preserve the sample integrity and the limited tilting range in liquid, the total electron dose of LP fast electron tomography was lower by one order of magnitude, with values of $71~\text{e}^-/\text{Å}^2$ and $787~\text{e}^-/\text{Å}^2$ for tilt series acquisition in liquid and in vacuum, respectively (Fig.~\ref{fig_2}e). Importantly, no significant changes in the interparticle distances between Au NPs within the colloidal clusters were observed after fast tilt series acquisition (Supplementary Fig.~\ref{SI_fig_3}a-f).

\subsection{Optimization of tilt series denoising and alignment for LP fast electron tomography} 
\begin{figure}[!htb]
    \centering
    \includegraphics[width=0.9\textwidth]{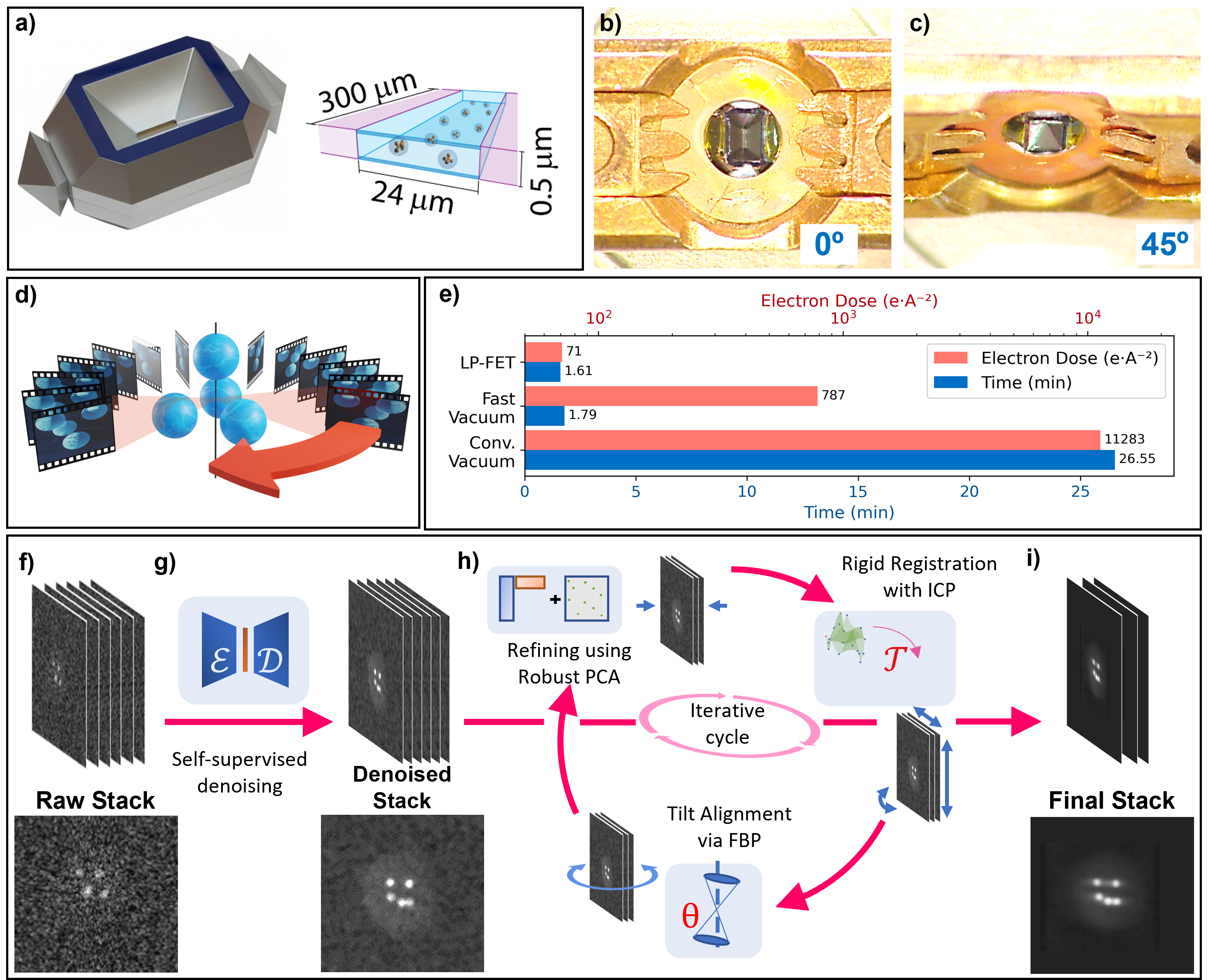}
    \caption{\textbf{Liquid-phase fast electron tomography}.
    a) Schematic illustration of a K-Kit LC used for experimental investigations, highlighting the LC dimensions.
    b-c) Optical micrographs of a K-Kit loaded on a single-tilt tomography holder, with, b) $0\degree$ and c) $45\degree$ tilting view, respectively.
    d) The LP fast electron tomography tilt series acquisition method continuously tilts the sample while recording projection images of the sample.
    e) Comparison of time and electron dose required for acquiring a complete tilt series using fast electron tomography in liquid and vacuum and conventional electron tomography in vacuum, highlighting the electron-beam efficiency of the fast electron tomography on liquid.
    f-i) fast electron tomography tilt series pre-processing workflow. 
    f) Representation of the raw tilt series stack with a sample image from the raw stack displayed.
    g) Illustration of the self-supervised denoising using CAE. A sample image from the denoised stack is displayed, demonstrating the effectiveness of the autoencoder denoising compared to the original one.
    h) Schematic overview of iterative process undertaken: refining the tilt series using RPCA, followed by rigid registration using the ICP method, and then tilt-axis alignment via FBP.
    i) The final processed stack, which is refined, aligned, and denoised, with a representative image displayed for clarity.}
    \label{fig_2}
\end{figure}

Mechanical movements of the goniometer during fast tilt series acquisition can lead to scanning distortions like streaking artifacts in the final 3D reconstruction, a phenomenon particularly evident in STEM mode\cite{vanrompay2021fast}. Factors such as low-dose imaging, inherent distortions, solvent presence, and the relatively thick $Si_xN_y$ window of the LC can adversely affect the SNR of the raw tilt series (Fig.~\ref{fig_2}f and Supplementary Movie 4).

In response to these challenges, we applied an advanced image processing and alignment approach (Fig.~\ref{fig_2}f-i; see the Methods section for details). We began with a self-supervised denoising technique utilizing convolutional autoencoders (CAE)\cite{gondara2016medical} (Fig.~\ref{fig_2}g; Supplementary Fig.~\ref{SI_fig_4}). This technique exploits the inherent sequential patterns present in tilt series images, which essentially are multiple representations of the object of interest from different angles. By utilizing this redundancy, the method effectively improves the SNR while retaining crucial structural details (Supplementary Fig.~\ref{SI_fig_5}). Following denoising, our iterative workflow consists of three stages to refine the tilt series. In the first stage, robust principal component analysis (RPCA)\cite{candes2011robust} (Fig.~\ref{fig_2}h) was applied to detect and eliminate distortions from the tilt series. At its core, RPCA decomposes the tilt series into two distinct matrices: a low-rank matrix and a sparse matrix. The low-rank matrix encapsulates the dominant, consistent features of the data, representing the underlying structure of the material. In contrast, the sparse matrix pinpoints irregularities or distortions, often arising from various sources during data acquisition. By isolating these anomalies, RPCA enhances the fidelity of the tilt series, facilitating improved registration, alignment, and 3D reconstruction. Next, we registered the tilt series projections using the iterative closest point (ICP) method, as depicted in Fig.~\ref{fig_2}h\cite{zhang2021fast}. The ICP method stands out from conventional algorithms due to its iterative approach, to minimize the difference between two clouds of points (computed from given tilt series and their low-rank RPCA component), making it particularly adept at handling the HAADF-STEM images. This iterative refinement ensures that even minor shifts or rotations that occur during image acquisition are accounted for. The final stage of each iteration focuses on aligning the tilt-axis for the tilt series. Accurate tilt-axis alignment is crucial because it ensures that the 3D reconstruction accurately represents the original structure without introducing artifacts. Misalignment can lead to distortions in the reconstructed volume, compromising the integrity of the analysis. This step involves comparing the tilt series to forward projections from an initial 3D reconstruction obtained using Filtered Back Projection (FBP), as shown in Fig.~\ref{fig_2}h\cite{houben2011refinement}. The effectiveness of this step stems from its ability to iteratively refine the alignment by leveraging the consistency in forward projections, ensuring that each subsequent iteration brings the tilt series closer to the true structural representation. This three-step procedure is repeated until sufficient convergence is obtained. At each iteration, the alignment of the tilt series images with their respective RPCA components is progressively refined, ensuring peak registration and alignment by the fifth cycle (Supplementary Fig.~\ref{SI_fig_6}). The outcome is a finely aligned tilt series, as illustrated in Fig.~\ref{fig_2}i and Supplementary Movie 5.

\subsubsection{Advanced 3D reconstruction algorithm for LP fast electron tomography}
After pre-processing to mitigate distortions, misalignments, and noise, as mentioned above, the challenge of the missing wedge due to a limited angular range needs to be addressed. Therefore, we devised an advanced 3D reconstruction algorithm. While conventional reconstruction algorithms, such as the simultaneous iterative reconstruction technique (SIRT)\cite{gilbert1972iterative}, maximum-likelihood expectation-maximization (ML-EM)\cite{moon1996expectation}, and total-variation minimization (TVM)\cite{goris2012electron}, fall short in addressing these challenges, the discrete algebraic reconstruction technique (DART)\cite{batenburg2011dart} has been widely adopted in electron tomography. DART capitalizes on the premise that materials are distinct and maintain constant intensity. Building on this foundation, we introduced an enhanced algorithm: Compressed-Sensing DART (CS-DART). CS-DART, an evolution of the standard DART, introduces additional refinements by integrating compressed sensing principles, thereby enhancing its ability to reconstruct images with greater accuracy, especially in scenarios with limited data, such as the missing wedge challenge in LP fast electron tomography. This method incorporates a shape smoothness prior, as detailed in the Methods section. Colloidal clusters, made up of Au NPs and polymeric shells, exhibit two distinct grey values in line with the DART reconstruction criteria. We hypothesize that these components exhibit smooth geometries, which play a crucial role in reducing the remaining noise and alignment errors during reconstruction (Supplementary Fig.~\ref{SI_fig_7}). Once the 3D reconstruction with CS-DART is complete, we quantify the structural characteristics of the NP-formed polyhedra using quantitative descriptors such as interparticle distance, surface area, volume, and regularity index (see Supplementary Notes for details).

\subsection{Quantitative 3D analysis for colloidal clusters in liquid phase}
\label{sec:results:AuNPColloidal}

\begin{figure}[!htb]
    \centering
    \includegraphics[width=0.9\textwidth]{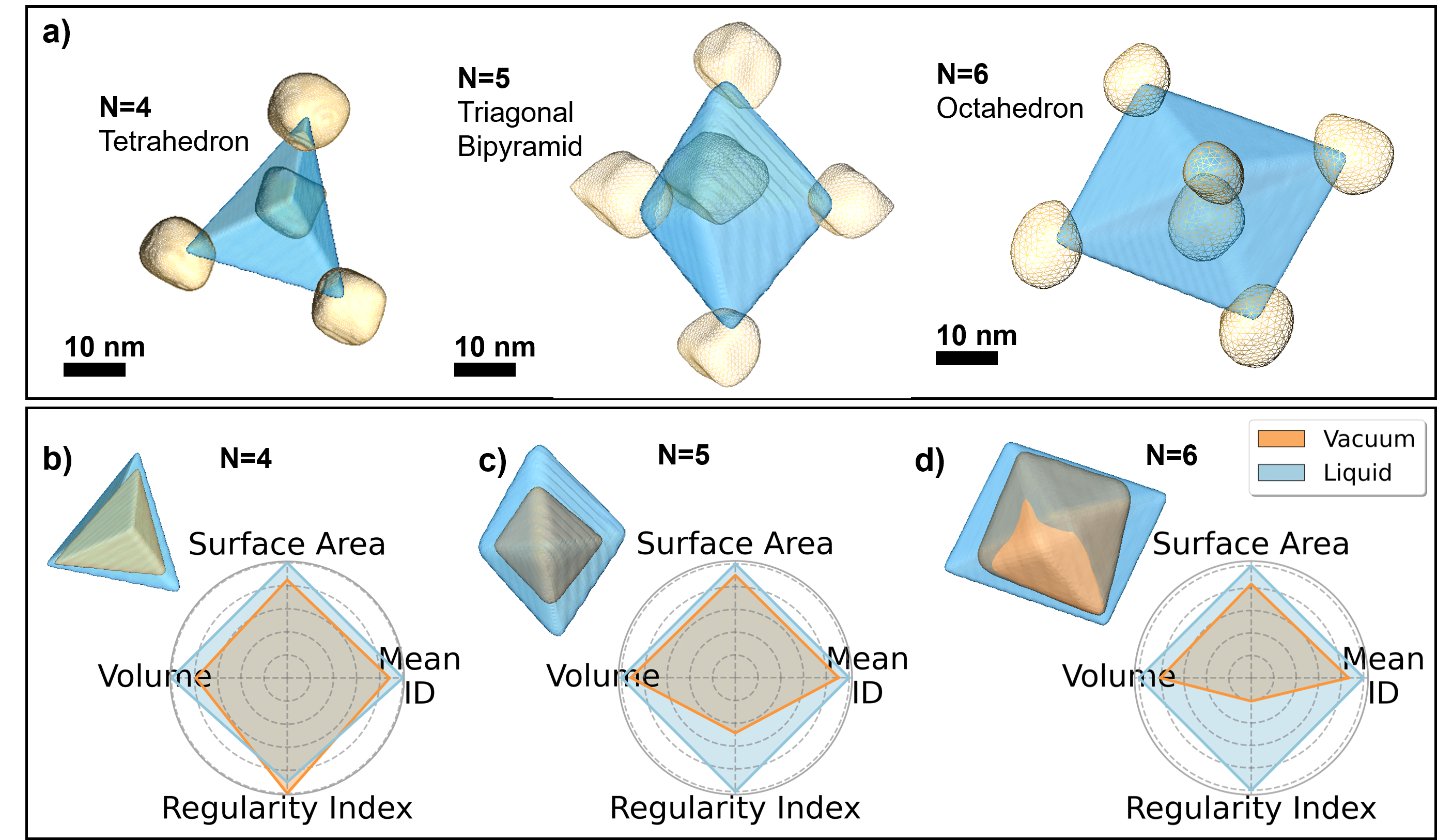}
    \caption{\textbf{Quantitative structural comparison between 3D reconstructions of colloidal clusters with different numbers of particles, implemented in liquid and vacuum conditions.}
    a) Polyhedra computed from the centroid positions of the Au NPs obtained through CS-DART reconstructions of three clusters containing N = 4, 5, 6 Au NPs in a liquid environment.
    b-d) Quantitative normalized structural comparison between polyhedra formed by the stacking of Au NP obtained from fast electron tomography in vacuum (depicted in orange) and liquid (depicted in blue), including mean interparticle distance (Mean ID), surface area, volume, and regularity index for b) N = 4, c) N = 5, and d) N = 6, respectively. Importantly, the Au NPs demonstrated a notable tendency to adopt regular but more condensed configurations when observed in a vacuum environment.}
    \label{fig_3}
\end{figure}

Our study primarily focused on the 3D arrangement of Au NPs within polymeric shells. Specifically, we observed that clusters containing 4, 5, and 6 Au NPs predominantly formed tetrahedral, trigonal bipyramidal, and octahedral structures, respectively (Fig.~\ref{fig_3}a; Supplementary Movies 6-8). However, reconstructing the polymeric shells proved challenging due to the low \textit{Z}-contrast arising from the water layer and the $Si_xN_y$ membrane of the LC chip.

To delve deeper into the structural nuances influenced by the environment, we compared the polyhedral structures obtained from fast electron tomography reconstructions in both vacuum and liquid settings. The radar charts in Fig.~\ref{fig_3}b-d visually contrast four quantitative descriptors (mean interparticle distance, volume, surface area, and regularity index) for the polyhedra formed in vacuum versus liquid across three assemblies (N = 4, 5, 6) (Table~\ref{table_1}; Supplementary Movies 6-8). Our quantitative analysis revealed that in a liquid environment, the average interparticle distance for clusters containing 4, 5, and 6 Au NPs was expanded by $13\%$, $10\%$, and $15\%$, respectively, compared to their vacuum counterparts. Similarly, the surface area and volume of these polyhedra in liquid were larger by varying percentages, indicating a more spacious arrangement of the Au NPs within the polymeric shells in liquid. 

To further understand the structural regularity, we introduce a regularity index, quantifying the similarity between an experimentally assembled 3D cluster formed by the Au NPs in different environments and an idealized, regular polyhedron (Table~\ref{table_1}). In a vacuum condition, the arrangement of Au NPs more closely resembles regular and compact polyhedra than in a liquid environment (Fig.~\ref{fig_3}). This suggests that capillary forces acting on the polymeric shells during drying compress the Au NPs into regular configurations. We observed a more pronounced 3D structural disparity between clusters in vacuum and liquid, with increasing numbers of Au NPs. This may be attributed to the tetrahedron (N = 4) having the highest packing fraction and thus, the least free volume among the studied polyhedra. As the number of Au NPs increases, they likely have increased mobility within the polymeric shell when dispersed in water. This disparity in structures between the liquid phase and vacuum is likely attributed to the presence of remaining solvent (THF) within the clusters when still in the liquid phase, which is removed during drying in vacuum, additionally leading to deformation of the polymeric shell by capillary forces during the same drying process prior to ET.

In conclusion, our findings emphasize the significant influence of the experimental environment on the structural characterization of colloidal clusters. We advocate for electron tomography in a liquid environment as it avoids capillary forces, offering a more authentic and representative 3D structural characterization.

\begin{table}[h!]
\centering
\setlength{\tabcolsep}{4pt}
{\small
\begin{tabular}{c| cc | cc | cc | cc}
\toprule
\textbf{N} & \multicolumn{2}{c|}{\textbf{mean ID (nm)}}  & \multicolumn{2}{c|}{\textbf{SA ($10^3$ nm$^2$)}} & \multicolumn{2}{c|}{\textbf{Volume ($10^3$ nm$^3$)}} & \multicolumn{2}{c}{\textbf{RI}} \\
 & \textbf{Vacuum} & \textbf{Liquid} & \textbf{Vacuum} & \textbf{Liquid}  & \textbf{Vacuum} & \textbf{Liquid}  & \textbf{Vacuum} & \textbf{Liquid} \\
\midrule
4 & 26.51 & 29.96 ($+13\%$) &  1.31 & 1.54 ($+15\%$) & 2.42 & 3.07 ($+21\%$)& 7.27 & 6.56 ($-10\%$) \\
5 & 29.21 & 32.16 ($+10\%$) &  2.17 & 2.41 ($+10\%$) & 6.23 & 6.79 ($+8\%$) & 5.56 & 11.46 ($+52\%$) \\
6 & 28.88 & 33.33 ($+15\%$) & 2.66 & 3.18 ($+16\%$) & 9.96 & 12.29 ($+19\%$) & 3.44 & 16.63 ($+79\%$) \\
\bottomrule
\end{tabular}
}
\caption{\textbf{Characteristics of Au NP-formed polyhedra in vacuum versus liquid conditions.} The table presents metrics, such as the mean interparticle distance (Mean ID), surface area (SA), volume, and regularity index (RI) of the polyhedrons derived from CS-DART reconstructions, highlighting the influence of imaging conditions on the reconstructed structures. The RI indicates the degree of regularity in the arrangement of the Au NPs, with a smaller value suggesting the arrangement is closer to a regular polyhedron.}
\label{table_1}
\end{table}

\subsection{Characterizing bilayer assemblies of Au NRs in liquid phase}
\label{sec:results:AuNRBilayer}

Apart from colloidal assemblies, NPs are often organized on solid substrates. A well-known example is the organization of Au NRs for exploiting their unique (and anisotropic) localized surface plasmon resonance (LSPR) properties, \eg, in sensing based on surface-enhanced Raman scattering (SERS)\cite{langer2019present}. It has been shown that the formation of ordered Au NR multilayers can lead to highly efficient SERS substrates\cite{solis2017optimization, hamon2015collective}. However, the excitation of individual \textit{vs.} collective LSPR modes in Au NR assemblies not only depends on their degree of organization but also (and very strongly) on the interparticle distances within the assembly. Again, usual practice involves TEM or SEM (rarely ET) on dry samples in vacuum, which is likely to affect both the structure and the interparticle distance. In this context, we decided to investigate bilayers of self-assembled CTAB-coated single crystalline Au NRs. For measurements in liquid, we used a Tomochip, \ie, a modified LC chip based on the monolithic K-Kit LC design\cite{das2022high, patent}. Although this LC has a smaller window gap of 100 nm, it allows achieving a significantly larger tilt range (up to $\pm 70\degree$; Supplementary Fig.~\ref{SI_fig_8}a-c). This configuration not only ensures a thinner liquid layer but also improves the SNR and angular coverage for tilt series in liquid conditions, minimizing the increase of the effective thickness at higher angles. We additionally dropcasted a dispersion of the same Au NRs on a TEM grid and dried it for structural analysis in vacuum.

\begin{figure}[!htb]
    \centering
    \includegraphics[width=0.9\textwidth]{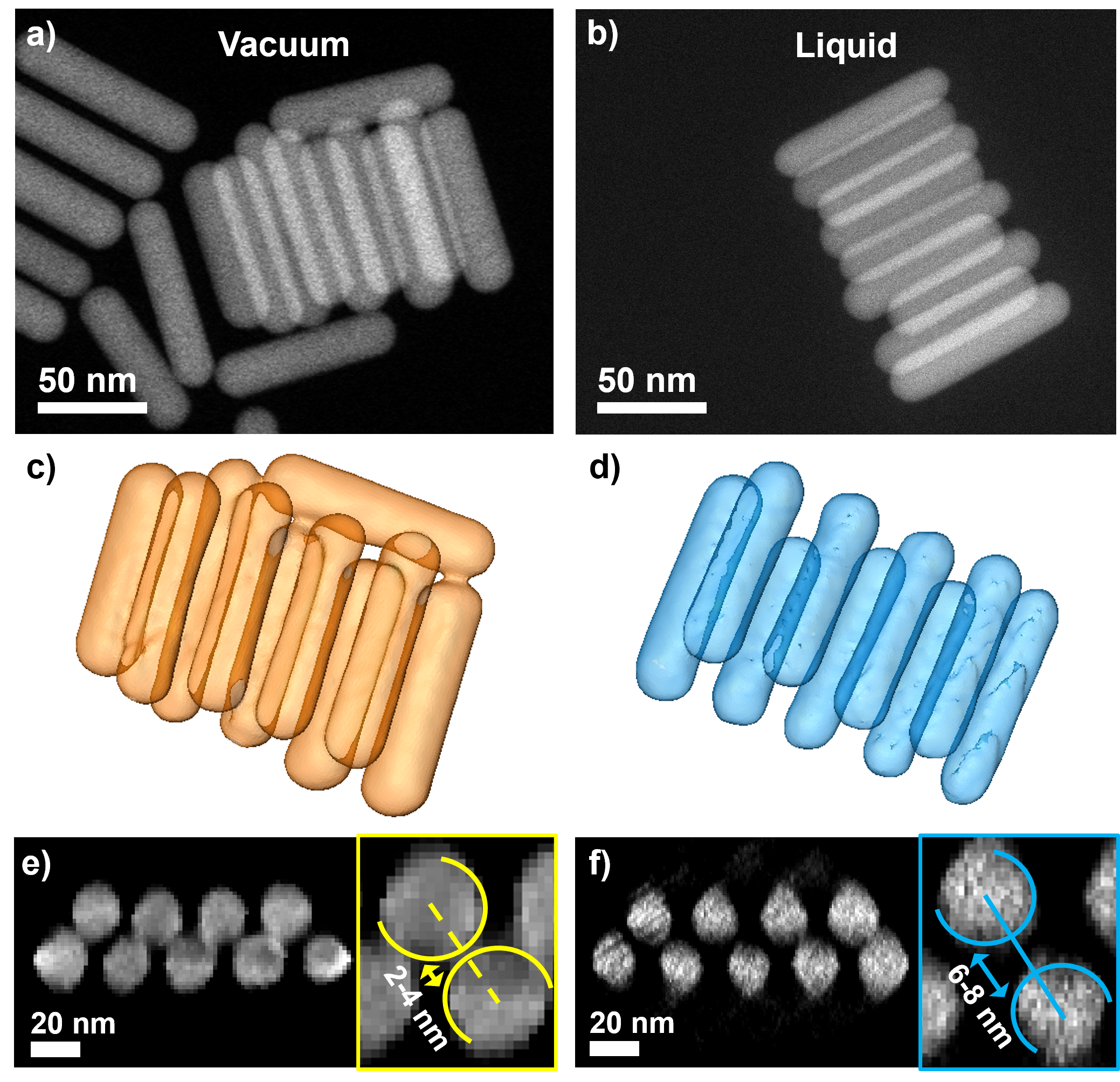}
    \caption{\textbf{Quantitative structural comparison of Au NR bilayer assemblies in vacuum and liquid-phase.}
    2D HAADF-STEM images of self-assembled Au NR superlattices measured in a) vacuum and b) liquid.
    c,d) 3D reconstructions and e,f) orthogonal views of superlattices assembled from two layers of Au NRs in c,e) vacuum and d,f) liquid.
    Insets of panels e) and f): zoomed-in views of two adjacent Au NRs, showing the disparity between surface-to-surface distances in vacuum and in liquid environment.
    Note that the transparency of the 3D renderings was increased for visual clarity.}
    \label{fig_4}
\end{figure}

We conducted fast electron tomography experiments in both vacuum and liquid environments, examining self-assembled bilayers of Au NRs, either dried on a TEM grid or encapsulated within the Tomochip (Fig.~\ref{fig_4}a,b; Supplementary Movie 9). The improved angular coverage and SNR facilitated the use of the conventional ML-EM algorithm\cite{moon1996expectation} to achieve precise 3D reconstructions in both settings. These reconstructions clearly allowed us to observe well-defined rod shapes for all Au NRs in either environment, organized in an AB-stacking pattern (Fig.~\ref{fig_4}c-f; Supplementary Movie 10). To further understand the environmental influence on Au NR stacking, we focused on the central region of the assemblies and measured the diameter of the Au NRs from 2D projections of the tilt series in vacuum and liquid, separately, ensuring our analysis was not affected by edge distortions (Supplementary Fig.~\ref{SI_fig_9}a-b). We then calculated the surface-to-surface distance by subtracting the radii of the Au NRs from the distance between their centers of mass (Supplementary Fig.~\ref{SI_fig_9}c-d, Supplementary Table~\ref{SI_table_1}). This mode of examination highlighted clear differences in surface-to-surface distances of both settings: 2-4 nm in vacuum \textit{vs.} 6-8 nm in liquid (Fig.~\ref{fig_4}e-f). The length of a fully stretched CTAB molecule is around 2.2 nm\cite{weidemaier1997photoinduced}. The surface-to-surface distance in liquid environment corresponds to almost four times the length of a CTAB molecule, where the two adjacent rods are expected to share a layer of counterions\cite{sau2005self}. We remark that, although the expected CTAB layer thickness on Au NRs has been characterized as approximately 3.2 nm\cite{gomez2012surfactant,meena2013understanding}, its precise structure — whether interdigitated bilayer or isolated micelles — is still under debate, in particular considering the influence of the experimental environment\cite{mosquera2023surfactant}. In contrast to the observation in liquid environment, the shorter surface-to-surface distance between NRs measured in vacuum indicates compression of the CTAB ligands. This is likely due to plastic deformation from capillary forces during sample drying, high vacuum conditions, or a combination of both. Our results highlight how LP fast electron tomography preserves the 3D structures of self-assembled Au NRs, avoiding distortions from capillary forces, in contrast with previous reports.

\clearpage


\section{Discussions}
\label{sec:discussions}
In this work, we have developed a methodology to characterize colloidal assemblies in 3D, in their native liquid environment. We addressed technical challenges, predominantly associated with commercial LCs, such as self-rotation and restricted tilt range. This was achieved through the use of advanced image processing algorithms, along with the development of the dedicated CS-DART 3D reconstruction algorithm. Moreover, we demonstrated that designing specialized LC devices specifically for ET can effectively mitigate the constraints of a limited tilt range. This approach has the potential to reduce the computational load required for precise data analysis and characterization. We successfully applied our approach to explore the 3D structural characterization of Au@PS colloidal clusters in their native liquid environment, contrasting these findings with structures observed in vacuum. Notably, we identified that capillary forces during the drying phase in vacuum conditions induced structural compression, leading to a more compact arrangement of Au NPs. Conversely, in liquid conditions, the NPs adopted a more expansive polyhedral shape. These structural variations can significantly influence the assemblies’ optoelectronic properties. Furthermore, we adapted our technique to study CTAB-stabilized Au NRs in bilayer self-assembly using a specialized Tomochip for 3D liquid characterization. In vacuum, the surface-to-surface distances of these bilayers are shorter, indicative of capillary effects. Yet, in a liquid setting, these distances align with both bulk measurements and theoretical predictions, emphasizing the importance of the experimental environment in electron microscopy studies.

Our study underscores the importance of analyzing colloidal assemblies in conditions that closely resemble their natural or application-specific environments. Analyzing them in non-native settings can introduce structural changes, potentially affecting our understanding of the assemblies’ true nature and properties. This information is crucial when adjusting these assemblies, especially in areas where specific changes, such as plasmonic effects, are desired. Moreover, we expect that our methodology can be applied to study the structure of a broad range of colloidal assemblies, giving insight into the structure formation mechanisms and structure-property relations of nanomaterials. While we have achieved notable progress, refining the structural analysis of colloidal assemblies remains an ongoing effort. Upcoming advancements, such as improved Tomochips and low-dose TEM techniques\cite{lazic2016phase,yu2022real}, aim to boost both precision and range in characterizations, proving essential for studying diverse assemblies, from varied sizes to those with low atomic numbers.


\section{Methods}
\label{sec:methods}

\subsection{Chemicals}
Gold (III) chloride hydrate (\ce{HAuCl4}, $\geq 99.9\%$), hexadecyltrimethylammonium bromide (CTAB, $\geq 99\%$), cetyltrimethylammonium chloride solution (CTAC, 25 wt. \% in \ce{H2O}), sodium borohydride (\ce{NaBH4}, $\geq 96\%$), L-ascorbic acid (AA, $\geq 99\%$), silver nitrate (\ce{AgNO3}, $\geq 99\%$), sodium oleate (\ce{NaOL}, $\geq 99\%$), hydrochloric acid (\ce{HCl}, $37\%$) and tetrahydrofuran (THF, $\geq 99\%$) were purchased from Sigma-Aldrich-Merck. Thiol-terminated polystyrene (PS509-SH, MW: 53K) and poly (styrene-b-acrylic acid) (PS403-b-PAA62) were purchased from Polymer Source. All chemicals were used without further purification. Milli-Q water (resistivity $18.2 \, \text{M}\Omega \cdot \text{cm}$ at $25^\circ \text{C}$) was used in all experiments. All glassware was washed with aqua regia, rinsed with Milli-Q water, and dried before use.

\subsection{Preparation of Au@PS colloidal clusters}
Gold seeds ($\sim 1.5~nm$) were prepared by fast reduction of \ce{HAuCl4} ($5~mL$, $0.25~mM$) with freshly prepared \ce{NaBH4} ($0.3~mL$, $10~mM$) in aqueous CTAB solution ($100~mM$), under vigorous stirring\cite{zheng2014successive}. The solution color changed from yellow to brownish yellow and the seed solution was aged at $27^\circ \text{C}$ for $30~min$ before use, to promote the decomposition of sodium borohydride. The seed was used for the growth of gold nanospheres and nanorods. An aliquot of gold seed solution ($0.6~mL$) was added under vigorous stirring to a growth solution containing CTAC ($100~mL$, $100~mM$), \ce{HAuCl4} ($0.36~mL$, $50~mM$), and ascorbic acid ($0.36~mL$, $100~mM$). The mixture was left undisturbed for 12 h at $25^\circ \text{C}$. Upon synthesis, the solution containing $12~nm$ Au NPs was centrifuged ($9000~rpm$, 2 h) to remove excess of CTAC and ascorbic acid, and finally redispersed in water to a final gold concentration of $5~mM$. The average diameter determined from TEM images was $12 \pm 1~nm$.
To replace the cationic surfactant with a hydrophobic polymer, thiolated polystyrene (PS-SH) was used. The colloidal dispersion containing $12~nm$ Au NPs ($2~mL$, $5~mM$) was added dropwise under sonication to a dispersion of PS-SH (1 molecule of PS-SH per $nm^2$ of Au surface) in THF ($20~mL$). The solution was left for 15 min in an ultrasonic bath. To ensure ligand exchange, the resulting mixture was left undisturbed for 12 h, and then centrifuged twice ($8000~rpm$, 30 min). The particles were finally dispersed in THF to a final gold concentration of $2.5~mM$.
The clustering of PS-functionalized Au NPs was carried out according to our previously reported\cite{sanchez2012hydrophobic}. An aliquot of water ($0.8~mL$) was added dropwise to the PS-functionalized Au NPs in THF ($3.2~mL$) under magnetic stirring. The final concentration of metallic gold in the mixture was $0.25~mM$. The solution was left undisturbed for 5 min, and then a solution of PS403-b-PAA62 in THF ($0.4~mL$, $6~mg/mL$) was added dropwise under magnetic stirring. Subsequently, the water content was increased up to $35~wt\%$, followed by increasing the temperature up to $70^\circ \text{C}$, which was maintained for 30 min. The clusters dispersion was centrifuged twice ($8500~rpm$, 30 min) and redispersed in water.

\subsection{Synthesis of Au NRs}
Au NRs were prepared through the seeded growth method, based on the reduction of \ce{HAuCl4} on CTAB-stabilized gold seeds in the presence of silver ions\cite{ye2013using}. To prepare the growth solution, $1.8~g$ of CTAB and $0.25~g$ of NaOL were dissolved in $100~mL$ of warm Milli-Q water ($50^\circ \text{C}$). Once sodium oleate was completely dissolved, the mixture was cooled down to $30^\circ \text{C}$ and AgNO3 ($4.8~mL$, $4~mM$) was added under stirring. The mixture was kept at $30^\circ \text{C}$ for 15 min after which HAuCl4 was added ($0.5~mL$, $100~mM$) under vigorous stirring. The mixture became colourless after 20 min at $30^\circ \text{C}$ and then HCl ($0.42~mL$, $37\%$) was introduced. After 15 min of stirring, AA ($0.25~mL$, $64~mM$) was added, and the solution was vigorously stirring for 30 seconds. Finally, a certain volume of seed solution ($0.16~mL$, $0.25~mM$) was injected into the growth solution under vigorous stirring for 5 minutes, and then the solution was left undisturbed al $30^\circ \text{C}$ for 12 h. An aliquot of the Au NR dispersion ($2~mL$) was centrifuged twice ($8000~rpm$, 30 min) to remove excess reactants, and then redispersed in a dilute CTAB aqueous solution ($0.25~mL$, $0.2~mM$). The length and width of the obtained NRs were $67 \pm 2~nm$ and $19 \pm 1~nm$, respectively.

\subsection{Electron microscopy sample preparation and measurements}
For a typical sample preparation for conventional and fast electron tomography measurement in vacuum, \SI{2} {\micro L} of colloidal dispersion was dropcast on a Quantifoil (2/2, 200 mesh) copper grid and was dried under an ambient environment. For a fast liquid electron tomography experiment on the Au@PS colloidal asseamblies, \SI{2} {\micro L} of the colloidal dispersion was loaded into commercial monolithic LC (K-Kit from Bio MA-TEK) with a window gap of \SI{0.5} {\micro m} under capillary forces, followed by being sealed with water-resistant glue. For a typical experiment on the Au NRs bilayer assemblies, the dispersion was loaded in a similar manner but using a Tomochip 0.1 {\micro m} window gap.

\subsection{Fast electron tomography in vacuum and liquid}
\label{tomo}
All tilt series were obtained from a ``cubed'' aberration-corrected Thermo Fisher Titan microscope microscope at room temperature with an acceleration voltage of 200 $kV$. A Fischione model 2020 single tilt holder was used for the fast acquisition of the tilt series both in vacuum and liquid. For fast electron tomography tilt series acquisitions in vacuum, a tilt range of $\pm72^{\circ}$ and $\pm74^{\circ}$ was applied for colloidal clusters ($N=4$, $N=5$ and $N=6$) and self-assembled Au NRs, respectively. For fast electron tomography tilt series acquisitions in liquid, a tilt range from $-48^{\circ}$ to $46^{\circ}$, $-48^{\circ}$ to $44^{\circ}$, $-46^{\circ}$ to $46^{\circ}$, and $-56^{\circ}$ to $58^{\circ}$ was applied for colloidal clusters of $N=4$, $N=5$, $N=6$, and self-assembled Au NRs, respectively. The total dose employed for each case was calculated by multiplying the dose per frame by the number of frames. Table 2 summarizes the experimental conditions from each of the tilt series acquired in this study.

\begin{table}[h!]
\centering
\setlength{\tabcolsep}{4pt}
{\resizebox{\columnwidth}{!}{\begin{tabular}{c c|c|c|c|c|c|c|c|c}
\toprule
    ~ & \textbf{Sample} & \shortstack{\textbf{Min.}\\\textbf{angle}} & \shortstack{\textbf{Max.}\\\textbf{angle}} & \shortstack{\textbf{Total tilt-}\\\textbf{series frames}} & \shortstack{\textbf{Current}\\\textbf{(pA)}} & \shortstack{\textbf{dwell time}\\\textbf{($\mu$s)}} & \shortstack{\textbf{pixel size}\\\textbf{(pm)}} & \shortstack{\textbf{dose per}\\\textbf{frame}\\\textbf{($\text{e}^-/\text{Å}^2$)}} & \shortstack{\textbf{Total dose}\\\textbf{($\text{e}^-/\text{Å}^2$)}} \\
    \midrule
    \multirow{4}{0.1\textwidth}{\textbf{Vacuum}} & N = 4 & -72 & 72 & 341 & 40 & 1 & 1040 & 2.31 & 787 \\
    & N = 5 & -72 & 72 & 341 & 40 & 1 & 1040 & 2.31 & 787 \\
    & N = 6 & -72 & 72 & 341 & 40 & 1 & 1040 & 2.31 & 787 \\
    & AuNRs & -74 & 74 & 242 & 50 & 1 & 736 & 5.76 & 1394 \\
    \midrule
    \multirow{4}{0.1\textwidth}{\textbf{Liquid}}  & N = 4 & -46 & 46 & 154 & 2 & 0.5 & 367 & 0.46 & 71 \\
    ~ & N = 5 & -48 & 44 & 156 & 2 & 0.5 & 367 & 0.46 & 72 \\
    ~ & N = 6 & -46 & 46 & 154 & 2 & 0.5 & 367 & 0.46 & 71 \\
     ~ & AuNRs & -56 & 58 & 163 & 5 & 0.5 & 367 & 1.16 & 189 \\
\bottomrule
\end{tabular}
}}
\caption{\textbf{Summary of vacuum and liquid fast electron tomography experimental conditions.}}
\label{table:exp}
\end{table}

\subsection{Tilt-series processing}
\label{sec:fast electron tomography tilt pro-rec}

\subsubsection{Denoising tilt-series images with convolutional autoencoders}
Imaging nanoparticles in liquid environments, especially under low-dose conditions with water surrounding the particles, often introduces significant noise. This noise can degrade the quality and accuracy of tilt series images. Recognizing that these images represent the same object from different angles, we can exploit inherent patterns for denoising. To enhance the SNR of these images, we employed a self-supervised denoising mechanism using CAE\cite{gondara2016medical}. CAEs, a specialized architecture in unsupervised machine learning, are designed to reconstruct their input data and consist of an encoder, which compresses the input into a latent representation, and a decoder, which reconstructs the input from this latent space. For our application, the CAE is specifically tailored to serve as a denoising tool, trained to generate noise-free images from their noisy counterparts. The mathematical representation of the denoising autoencoder is:
\begin{align}
\text{Encoder: } \qquad \vz &= \mathcal{E}_{\vlambda} ( \vy ), \\
\text{Decoder: } \qquad \vy' &= \mathcal{D}_{\vgamma} ( \vz ), \\
\text{Loss Function: } \qquad \mathcal{L} (\vlambda, \vgamma) &= -\sum_{i} y_i \log(y'_i) + (1 - y_i) \log(1 - y'_i),
\end{align}
where $\vy \in \R^n$ denotes the noisy input image, $\mathcal{E}_{\vlambda}$ represents the encoder function parameterized by $\vlambda$, $\mathcal{D}_{\vgamma}$ is the decoder function parameterized by $\vgamma$, and $\mathcal{L}$ is the negative log likelihood, tailored to be robust against Poisson noise. Our CAE architecture was implemented using the PyTorch library as shown in Supplementary Fig.~\ref{SI_fig_4}. Training was conducted over 50 epochs with a batch size of 32, using the Adam optimizer with a learning rate of $10^{-4}$. By harnessing the patterns present in sequential tilt series images, the CAE effectively reconstructs noise-free versions. This approach not only elevates the SNR but also preserves the crucial structural details inherent to the images. In a comparative analysis against traditional denoising techniques, such as Gaussian smoothing, our methodology showcased a superiority in both noise attenuation and preservation of structural intricacies (see Supplementary Fig.~\ref{SI_fig_5}). 

\subsubsection{Refining tilt series using robust principal component analysis}
High-quality tilt series images are crucial for accurate 3D reconstructions. However, the intricacies of fast image acquisition can introduce distortions and anomalies that compromise subsequent analyses. While denoising techniques, such as the CAE we employed, are effective in enhancing the SNR, they primarily target random noise. Systematic distortions, outliers, or structured anomalies, which can arise due to various factors in the imaging process, may still persist post-denoising. These structured anomalies can have a pronounced impact on the accuracy of 3D reconstructions. To specifically address and rectify these structured distortions, we further refined our tilt series using RPCA\cite{candes2011robust}. RPCA, an advanced extension of classical Principal Component Analysis (PCA), offers a robust approach to data decomposition. While PCA assumes the observed data is a mix of low-rank components and Gaussian noise, RPCA decomposes a data matrix $\mY \in \R^{n \times a}$ (with $a$ being number of images) into a low-rank matrix $\mL \in \R^{n \times a}$ and a sparse matrix $\mS \in \R^{n \times a}$:
\begin{equation}
\underset{\mL, \mS}{\mbox{minimize}}  \quad \Vert \mL \Vert_* + \lambda \Vert \mS \Vert_1 \qquad \text{subject to} \qquad \mL + \mS = \mY.
\end{equation}
Here, $\Vert \cdot \Vert_*$ denotes the nuclear norm, approximating the rank of $\mL$. $\Vert \cdot \Vert_1$ is the $\ell_1$ norm, a convex approximation for the count of non-zero elements in $\mS$.
For our implementation, we utilized the augmented Lagrange multiplier method for optimization, a popular approach for RPCA. The algorithm was run for a maximum of $100$ iterations with a convergence criterion set at $10^{-4}$. The parameter $\lambda$ was empirically set at $1/\sqrt{\max(a,n)}$, where $a$ and $n$ are the dimensions of $\mY$, ensuring a balance between the low-rank and sparse components. Applying RPCA to the tilt series allows us to discern the primary patterns within the series, highlighting important structures and relationships between individual images. We then establish a threshold based on a specific percentile of the distribution of projection scores from RPCA. Images deviating significantly from this threshold are considered outliers with potential distortions. These outliers are excluded from the tilt series, ensuring a refined dataset ready for further processing.

\subsubsection{Image alignment using iterative closest point method}
Accurate 3D reconstructions from tilt series critically depend on the precise alignment of the tilt series images. Given that particles can rotate in the liquid during acquisition, it’s imperative to account for both translational and rotational misalignments. Merely Only using cross-correlation to register shifts is insufficient, as it predominantly addresses translational misalignments. Therefore, a more comprehensive approach, like rigid registration, becomes indispensable to ensure that each image in the series is aligned to a reference (in this case, we use low-rank component $\mL$ obtained from RPCA). The ICP method offers a robust solution for this challenge\cite{zhang2021fast}. Originally designed for 3D point cloud alignments, ICP’s iterative approach is well-suited for tilt series image registration. At each iteration, the algorithm identifies pairs of closest points between two images. These points are derived from prominent features within the images, extracted using Speeded-Up Robust Features (SURF). The optimal transformation (rotation and translation) is then calculated to minimize the distance between these point pairs. This transformation progressively aligns the images. The iterations continue until the algorithm converges to a minimal distance between corresponding points in the datasets or until a predefined number of iterations (typically set to 100) is reached. The mathematical objective of ICP is:
\begin{equation}
\underset{\mR, \vt }{\mbox{minimize}} \quad \sum_i \Vert \vp_i - (\mR \vq_i + \vt) \Vert^2,
\end{equation}
where $\mR \in \R^{2 \times 2}$ is the rotation matrix, $\vt \in \R^2$ is the translation vector, and $\vp_i \in \R^2$ and $\vq_i \in \R^2$ are the corresponding points from the two images being aligned. For the optimization, we employed the proximal gradient algorithm, which is adept at handling non-linear least squares problems. We introduced constraints on the shifts and rotations to ensure physically meaningful alignments. These constraints were set based on prior knowledge of the maximum possible misalignments during the imaging process. Rigid registration, as facilitated by ICP, is essential because it compensates for any minor shifts or rotations that might occur during image acquisition. This ensures that the images are consistently oriented and overlaid with precision, which is fundamental for generating coherent and accurate 3D reconstructions.

\subsubsection{Alignment of tilt-axis using filtered back projection}
Tilt-axis alignment is a critical step for accurate three-dimensional (3D) reconstruction from electron tomography tilt series. Even minor misalignments can lead to artifacts, diminishing the fidelity of the reconstructed 3D structures. Therefore, achieving precise tilt-axis alignment is imperative for obtaining reliable 3D reconstruction. To facilitate this, we employ the Filtered Back Projection (FBP) method. FBP effectively reconstructs an object from its projections by utilizing the Radon transform and its inverse. This process generates a rapid 3D reference model. For the tilt-axis alignment, our approach begins with creating forward projections from the 3D model obtained via FBP. These projections are then quantitatively compared with the original tilt series images. The comparison employs a similarity metric, specifically the Structured Similarity Index Measure (SSIM), to identify any misalignments. Identifying discrepancies between the FBP-generated projections and the original tilt series is crucial. These discrepancies indicate the extent and nature of misalignments. To rectify these misalignments, we use an algorithmic refinement process\cite{houben2011refinement}. In each iteration of this process, the tilt-axis orientation is adjusted based on the observed discrepancies from the preceding iteration. This iterative refinement employs a gradient descent optimization strategy, systematically reducing the discrepancy metric to reach an optimal alignment. The number of iterations for convergence is typically around 10, although this can vary depending on the initial degree of misalignment and the quality of the tilt images. Upon convergence, the tilt-axis is accurately aligned.

\subsubsection{Iterative workflow}
In the pre-processing of electron tomography tilt series, adopting an iterative workflow is essential due to the complex nature of image acquisition and the possibility of sample movements during the process. Our methodology encompasses three fundamental steps: distortion correction using RPCA, image alignment through ICP algorithm, and precise tilt-axis alignment. The accuracy of each step is critical because it significantly affects the subsequent stages. Initially, the RPCA method identifies and corrects distortions, producing a more accurate tilt series for the next phase. Subsequently, the ICP algorithm aligns the images, and this alignment is further refined through tilt-axis adjustment. This sequence of steps is repeated across five iterations for optimal results. The decision to limit the process to five iterations stems from empirical observations. Specifically, we monitor a similarity metric that compares the tilt series with the low-rank component $\mL$ (referenced in Supplementary Figure~\ref{SI_fig_6}). As the iterations proceed, this metric generally shows improvement, signaling better alignment and registration quality. However, after the fifth iteration, we observe a plateau in this metric, implying that additional iterations would yield minimal further improvements. This plateau indicates the point where the balance between computational effort and alignment quality is optimized, thereby justifying the choice of five iterations in our process.

\subsection{Advanced reconstruction method from tilt series} \label{sec:CS-DART}
To ensure consistency and avoid potential bias, the tilt series in vacuum and liquid was processed using the methodology outlined above. This involved denoising, alignment, and distortion removal. Notably, given the broad tilting angular range of $140\degree$ achieved in vacuum, the ML-EM algorithm implemented using the ASTRA Toolbox\cite{astra} was employed for reconstruction. Subsequent 3D visualizations were rendered using Amira 5.4.0. 

In the context of electron tomography, particularly for samples imaged in liquid, challenges arise due to the restricted tilting range, often limited to about $90\degree$. This limitation poses significant challenges even for advanced reconstruction algorithms like DART\cite{batenburg2011dart}. DART can struggle with incomplete datasets, as a limited tilt range often fails to provide comprehensive angular coverage, leading to ambiguities and potential inaccuracies in the reconstructed 3D volume. To address these challenges, we have developed the Compressed Sensing Discrete Algebraic Reconstruction Technique (CS-DART). This advanced algorithm enhances the conventional DART by incorporating a shape smoothness prior. The key innovation in CS-DART lies in its use of level-set methods for material discretization, representing material intensities through a combination of level-set functions. By integrating smoothness into the material surfaces and leveraging discrete cosine transform (DCT) basis for regularizing the reconstruction problem, CS-DART effectively overcomes the limitations posed by incomplete angular coverage. The mathematical formulation of CS-DART is as follows:
\begin{equation}
    \begin{split}
        \underset{\valpha}{\mbox{minimize}} \quad  \| \mW \vx ( \phi( \valpha) ) - \vy \|_2^2 , \qquad \mbox{subject to} \quad \| \valpha \|_1 \leq \tau,
    \end{split}
\end{equation}
where $\mW$ is the tomographic operator that discretizes the Radon transform, $\valpha$ are the DCT coefficients of the object under reconstruction, $\vy$ is the acquired, processed tilt series, $n$ are the number of voxels, and $\tau$ is the regularization parameter. Here, we have modeled the intensity as
\begin{equation}
    \vx(\vphi) = c_{\text{soft}} H(\vphi - c_{\text{soft}}) + c_{\text{hard}} H(\vphi - c_{\text{hard}}), \qquad \vphi = \mPsi \valpha,
\end{equation}
where $c_{\text{soft}}$ and $c_{\text{hard}}$ are the intensities of soft (polymeric) and hard (Au NPs) material, $H$ is the Heaviside function ensuring the imposition of discreteness, $\vphi$ is the discretized level-set function, and $\mPsi$ is a DCT basis. This condensed approach forms the basis of our method, which we have termed the Compressed Sensing Discrete Algebraic Reconstruction Technique (CS-DART). CS-DART utilizes the DCT basis to efficiently compress shape information in level-set functions, allowing for effective handling of large datasets by focusing on key shape features. This approach not only enhances reconstruction accuracy but also significantly reduces computational complexity, as the number of functions in the basis is much lower than the voxel count. By capturing the essential features of the sample and minimizing redundancy, CS-DART streamlines the reconstruction of complex structures from incomplete datasets. The optimization problem is solved iteratively via the well-known Fast Iterative Shrinkage-Thresholding Algorithm\cite{beck2009fast}. This method efficiently accommodates the non-differentiability introduced by the $\ell_1$-norm in our objective function. Additionally, we use an approximation of the Heaviside function to compute the gradient of the loss function\cite{kadu2016salt}. In each iteration, the DCT coefficients, $\valpha$, are updated based on the gradient of the data fidelity term\cite{kadu2017parametric}. Moreover, the Radon transform was implemented using ASTRA Toolbox\cite{astra}. CS-DART algorithm accurately reconstructs heterogeneous structures from incomplete datasets, making it particularly beneficial for analyzing these colloidal assemblies under limited tilt ranges.

We remark that our advanced reconstruction algorithm recreates the morphology of the assembly while overcoming the challenges posed by missing wedge artifacts and noise. A comparison of the 3D reconstructions obtained using our proposed method and conventional reconstruction methods is presented in Supplementary Fig.~\ref{SI_fig_7}. Our results underline the enhanced performance and accuracy of our proposed CS-DART method, especially in preserving the fine structural details of the colloidal assemblies.

Due to enhanced angular sampling and improved SNR realized by Tomochip ($130\degree$ angular range), bilayer assemblies of Au NRs were reconstructed using ML-EM algorithm implemented in ASTRA Toolbox\cite{astra}.

\section*{Acknowledgement}

S.B., D.A.E., D.W., N.O., and A.K. acknowledge financial support from ERC Consolidator Grant Number 815128 REALNANO and Horizon Europe MSCA-SE no. 101131111 – DELIGHT. D.W. acknowledges an Individual Fellowship funded by the Marie-Sklodowska-Curie Actions (MSCA) in Horizon 2020 program (grant 894254 SuprAtom). L.M.L.M. acknowledges financial support from Project PID2020-117779RB-I00, State Research Agency of Spain, Ministry of Science and Innovation.


\clearpage

\makeatletter
\renewcommand \thesection{S\@arabic\c@section}
\renewcommand\thetable{S\@arabic\c@table}
\renewcommand \thefigure{S\@arabic\c@figure}
\makeatother

\setcounter{section}{0}
\section*{Supplementary Notes}
\section{Electron beam damage comparison on colloidal assemblies: fast \textit{vs.} conventional acquisition methodologies}
To better understand the impact of electron beam exposure on colloidal assemblies, we compared different acquisition methodologies, specifically fast vs. conventional methods. This comparison is crucial, as prolonged exposure to the electron beam could potentially alter the structure of the studied samples. Presented in Supplementary Fig. \ref{SI_fig_1}, HAADF-STEM images of polymeric assemblies with different structures ($N=4, 5, 6$), before and after both acquisition approaches, are presented. It is interesting to note the difference in the size of the assemblies post-acquisition. Whereas the fast acquisition method seems to preserve the original size with minimal alterations, the conventional methodology, on the other hand, demonstrates evident shrinkage, by 2 - 3 nm on average. These findings confirm the importance of carefully selecting acquisition methods, depending on the sensitivity of the samples under investigation\cite{vanrompay2021fast}.

\section{Dynamic behavior of colloidal assemblies under flow in a liquid phase TEM holder}
To gain a deeper understanding of colloidal assembly dynamics, especially under flow conditions, we employed a liquid phase TEM holder to capture the real-time behavior of these assemblies. Supplementary Fig. \ref{SI_fig_2}a shows the flow-induced liquid phase setup (Stream from  DENSsolutions) utilized to observe the colloidal clusters. As observed in the subsequent HAADF-STEM image series (Supplementary Fig. \ref{SI_fig_2}b-d), distinct colloidal clusters demonstrated significant translational and rotational motion over time. Examples of the dynamic behavior are indicated by the dashed yellow circle. However, a critical observation was the occurrence of electron beam damage after 33 seconds of exposure. This phenomenon resulted in unintended aggregation of the colloidal clusters. A more detailed visual representation is presented in Supplementary Movie 3.

\section{Quantitative Indicators}
\label{sec:QuantDescriptors}
Regarding 3D tomographic imaging of colloidal assemblies of Au nanoparticles (NPs) within a polymeric shell, surface area, volume, and regularity index serve as vital metrics for characterizing assembly structures. By analyzing the polyhedron formed by the assembled Au NPs inside the polymeric shell, these metrics are calculated.

Surface area measures the combined area of the polyhedron’s faces, providing insight into the extent of surface interactions between NPs and the polymeric shell, which can influence the stability of the colloidal assembly. To calculate surface area, the areas of all the polyhedron’s faces are summed up:

\setcounter{equation}{0} 
\begin{equation}
    S = \sum A_i,
\end{equation}
where $A_i$ is the area of the $i^{\text{th}}$ face of the polyhedron.

The volume quantifies the space enclosed by the polyhedron formed by the Au NPs. This metric offers information about the density and packing of the NPs within the polymeric shell, which can affect the assembly’s mechanical, optical, and electronic properties. We calculate the volume ($V$) by dividing the polyhedron into smaller parts, such as tetrahedrons or cubes, and summing their volumes:
\begin{equation}
    V = \sum V_i,
\end{equation}
where $V_i$ is the volume of the $i^{\text{th}}$ tetrahedron or cube.

The regularity index measures the polyhedron’s regularity. This metric provides information about the degree of order and symmetry in the assembly, which can influence physical and chemical properties of the colloidal assembly. We calculate the regularity index ($R$) by comparing the polyhedron to a regular polyhedron with the same number of faces and vertices and measuring the deviation from the regular shape in terms of the angles and lengths of the edges and faces:
\begin{equation}
R = \frac{{\sum_i ((\theta_i-\theta_r)^2+(l_i-l_r)^2)/n}}{{\sum_i (d_i^2/n)}},
\end{equation}
where $\theta_i$ and $\theta_r$ are the angles of the $i^{\text{th}}$ and regular faces of the polyhedron, $l_i$ and $l_r$ are the lengths of the $i^{\text{th}}$ and regular edges of the polyhedron, $d_i$ is the distance of the $i^{\text{th}}$ vertex from the center of the polyhedron, and $n$ is the number of faces of the polyhedron. For reference, the ideal or `regular’ polyhedron considered has identical face and vertex counts as the Au NP-formed polyhedron. The closer the value of $R$ is to zero, the more regular or symmetric the polyhedron. A lower RI indicates a structure closely resembling an ideal polyhedron, suggesting high order and symmetry. Conversely, a higher RI points to irregularity.

\section{Determining Au NRs assembly surface-to-surface distances}
\label{Rods_RGB}
To understand the assembly patterns of Au NRs, we examined the surface-to-surface distances under both dry and liquid conditions. As evident from the HAADF-STEM projection images (Supplementary Fig. \ref{SI_fig_9}), distinct assembly configurations were observed for both vacuum and liquid environments. These images not only allowed us to visualize the assembly patterns but also facilitated accurate identification and measurement of individual rod diameters. To obtain accurate measurements, we took an orthogonal projection from the 3D reconstruction that was directly facing the rods. Using MATLAB’s `imfindcircles’ function, we pinpointed the center and diameter of each rod. With this data, we calculated the surface-to-surface distance between rods by measuring the direct distance between their centers and adjusting for their size. The surface-to-surface distance between the different rods (R, G, B) was tabulated and compared under dry and liquid conditions, as detailed in Supplementary Table\ref{SI_table_1}. The table showcases variations in the distances between the rods under the two conditions, providing essential insights into the role of environment on the self-assembly behavior of NPs.

\clearpage

\section*{Supplementary Figures}

\setcounter{figure}{0}

\begin{figure}[!htb]
    \centering
    \includegraphics[width=0.9\textwidth]{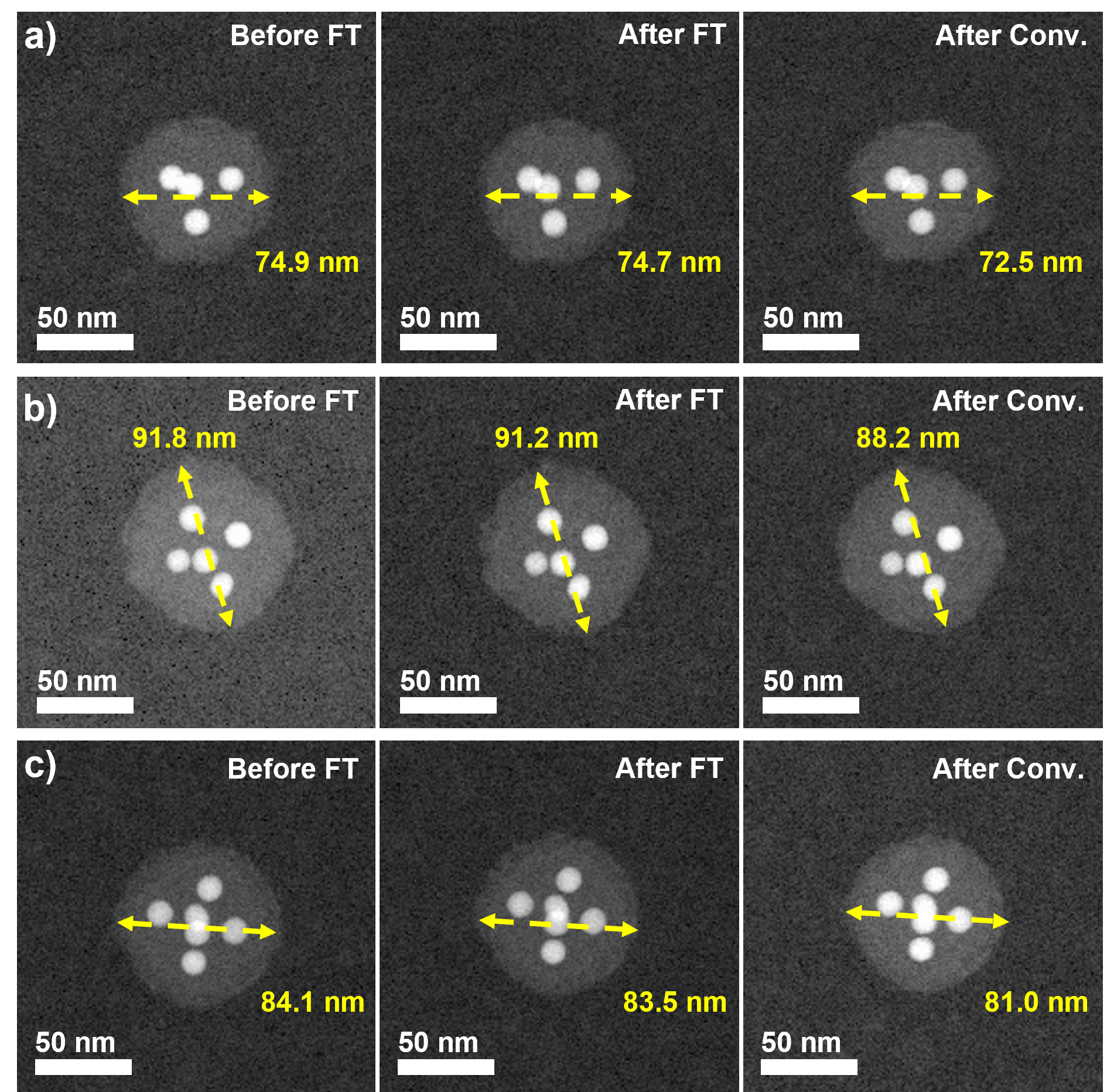}
    \caption{\textbf{Comparison of electron beam damage on colloidal clusters.} HAADF-STEM images acquired before, after fast acquisition, and after conventional tomography tilt series acquisition, for colloidal assemblies with different numbers of Au NPs, a) $N = 4$, b) $N = 5$ and c) $N = 6$. It can be observed that fast acquisition shows a minimal size change, but conventional acquisition shows a shrinkage of 2 - 3 nm.}
    \label{SI_fig_1}
\end{figure}
\clearpage

\begin{figure}[!htb]
    \centering
    \includegraphics[width=0.9\textwidth]{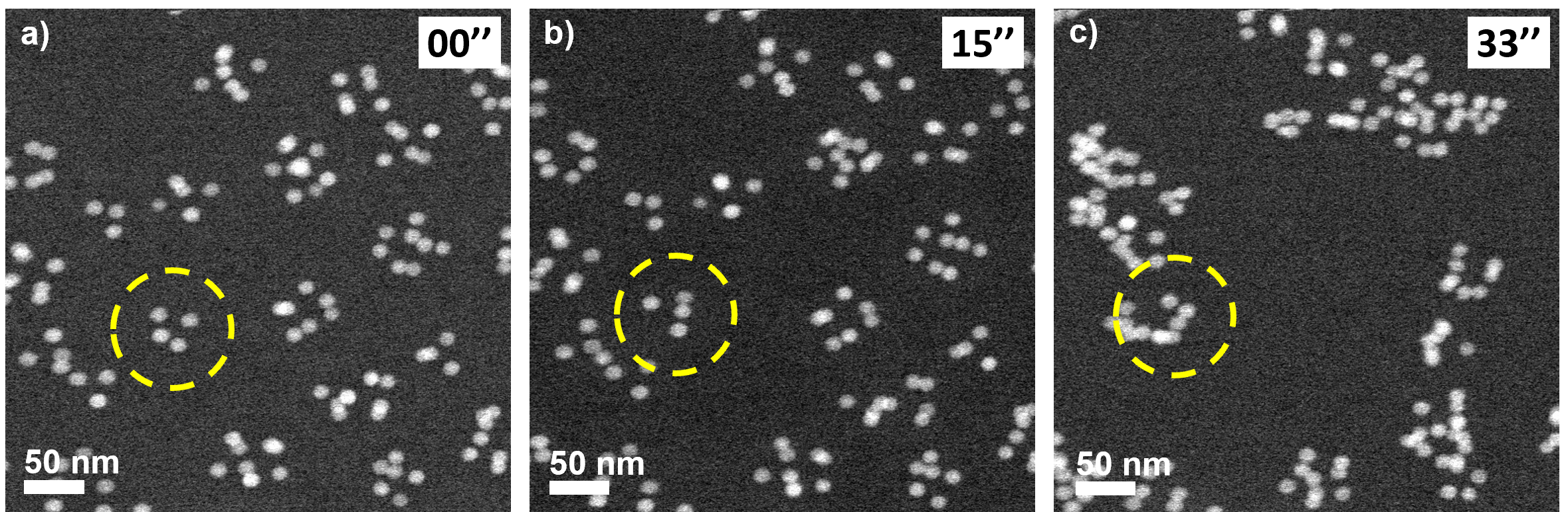}
    \caption{\textbf{Challenges in 3D characterization of colloidal clusters by tracking translational and rotational motions.} 
    Series of HAADF-STEM images of clusters obtained in a commercial \ce{Si3N4} LC chamber under flowing at different time lapses: a) 0 seconds, b) 15 seconds, c) 33 seconds.
    Some colloidal clusters were observed to translate and rotate freely, as denoted by the dashed yellow circle. However, electron-beam damage is evident, even after only 33 seconds of exposure time, leading to aggregation of the colloidal clusters.
    Note that the presence of a liquid phase and thick \ce{Si3N4} LC windows resulted in the polymeric shells being invisible. The dynamic process can be better appreciated in Supplementary Movie 3.}
    \label{SI_fig_2}
\end{figure}
\clearpage

\begin{figure}[!htb]
    \centering
    \includegraphics[width=0.9\textwidth]{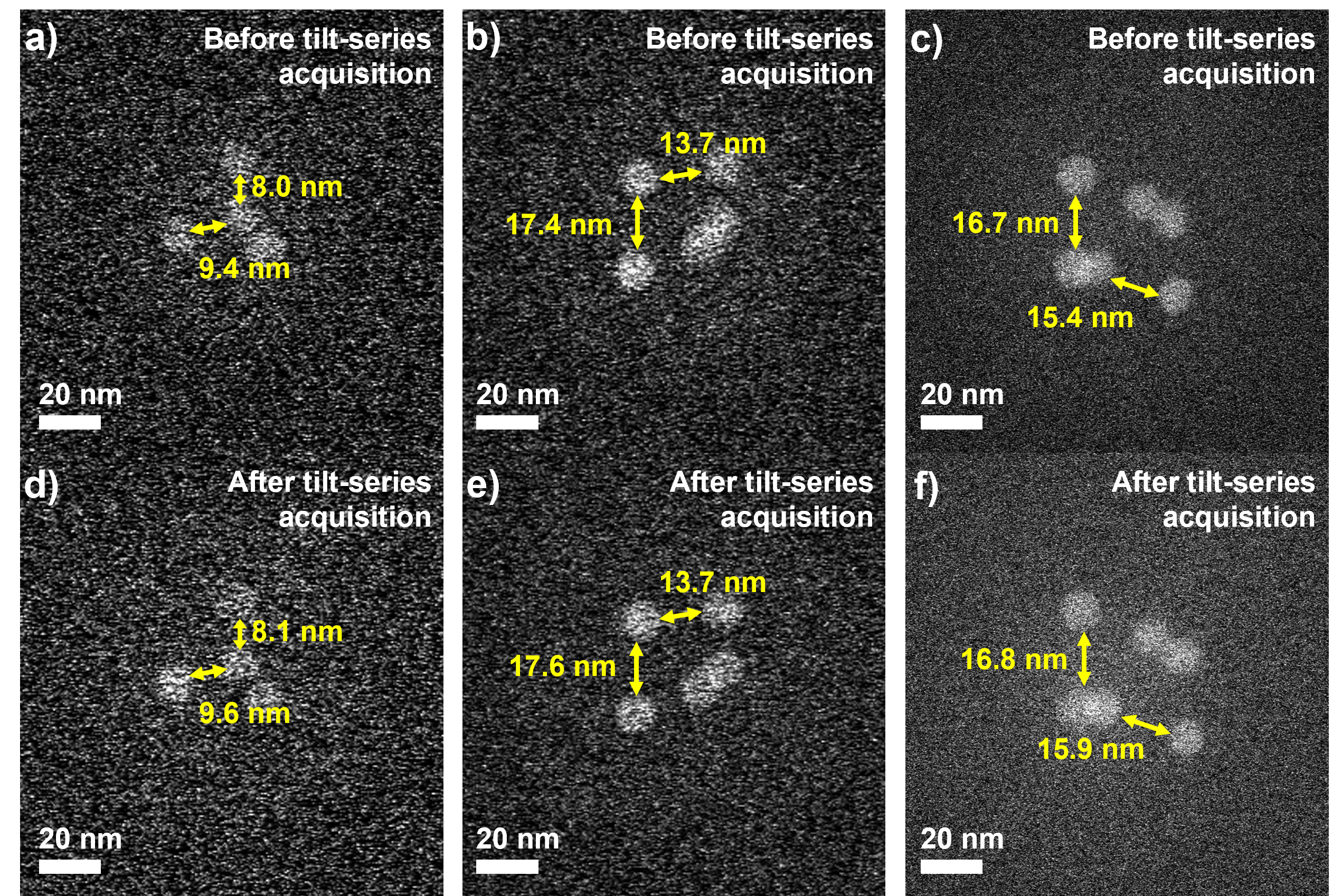}
    \caption{\textbf{Comparison of the structure, before and after fast tilt series acquisition in liquid.} HAADF-STEM images of different assemblies investigated in liquid environment, acquired before (a, b and c) and after (d, e and f) the tilt series acquisition for N = 4, 5 and 6 respectively. No significant structure difference is observed after tilt series acquisition.}
    \label{SI_fig_3}
\end{figure}
\clearpage

\begin{figure}[!htb]
    \centering
    \includegraphics[width=0.9\textwidth]{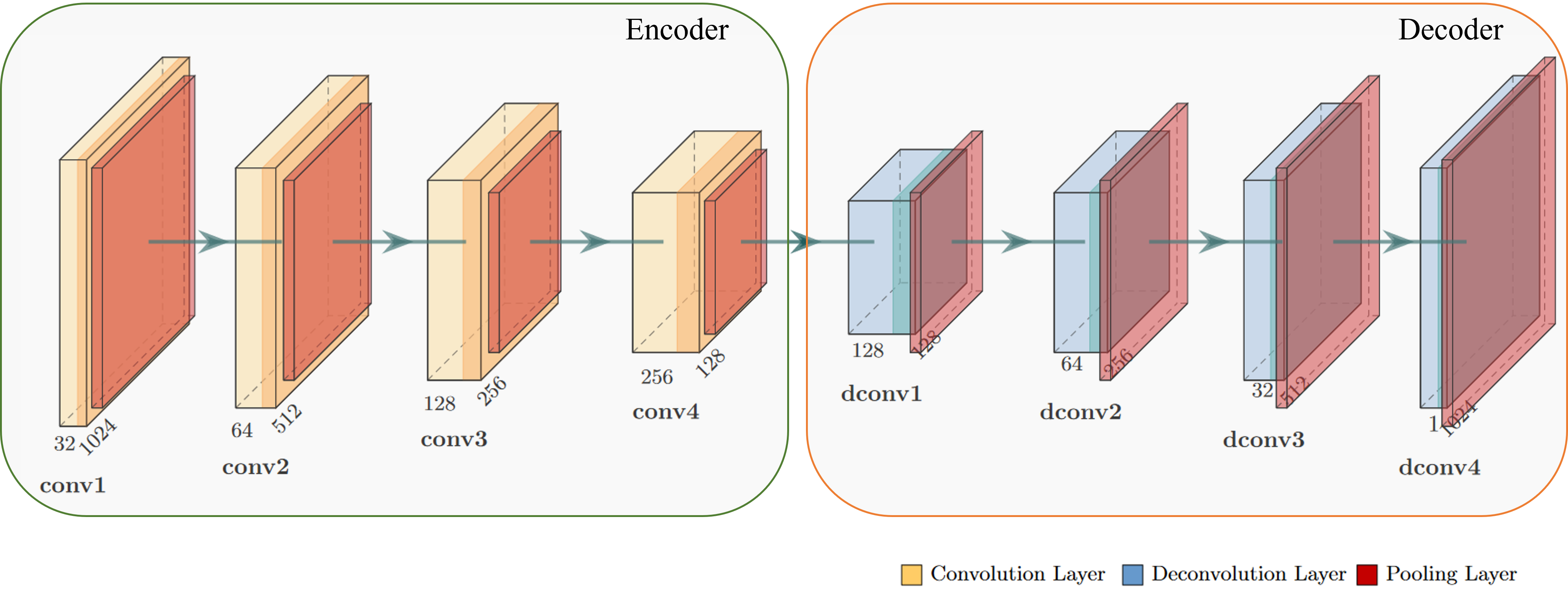}
    \caption{\textbf{Schematic representation of a Convolutional Autoencoders (CAE) architecture.} The CAE consists of an encoder and a decoder. The encoder progressively reduces the spatial dimensions of the input image through convolutional and pooling layers, capturing its salient features in a compressed latent space. The decoder then reconstructs the original image from this compressed representation by using deconvolutional layers and upsampling. The input to the CAE is a grayscale image of size $1024 \times 1024$, and the output is a reconstructed (denoised) image of the same size.}
    \label{SI_fig_4}
\end{figure}
\clearpage

\begin{figure}[!htb]
    \centering
    \includegraphics[width=0.9\textwidth]{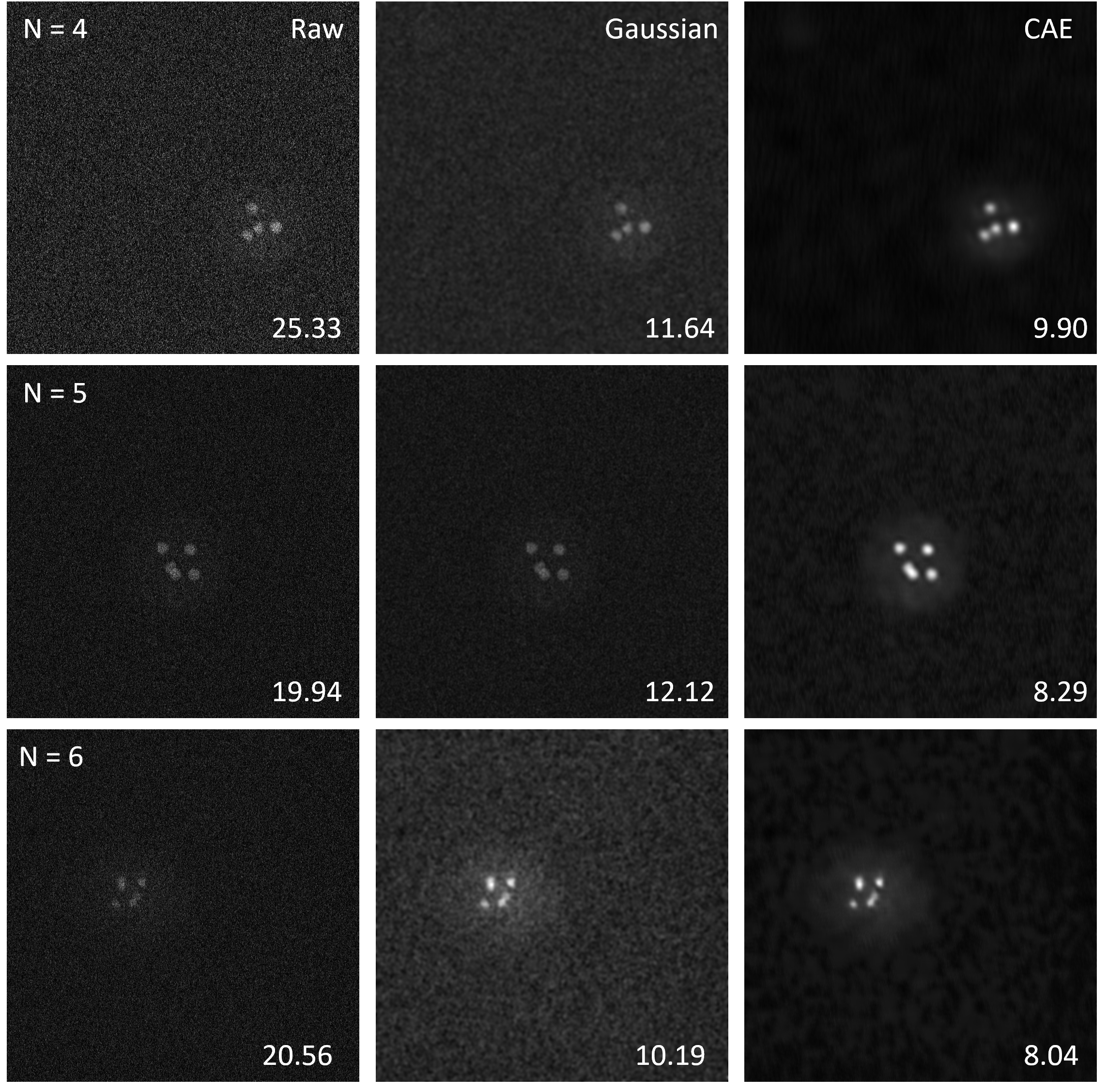}
    \caption{\textbf{A side-by-side comparison of two image denoising techniques.} The leftmost column presents the original images, the center column displays the outcomes of Gaussian denoising, and the rightmost column reveals the results using CAE denoising. Each row represents a different colloidal assembly, all imaged at a $0^\circ$ tilt-angle. The quality of denoising is quantified using the Naturalness Image Quality Evaluator (NIQE) score, displayed at the bottom right of each image. A lower NIQE score indicates superior image clarity and denoising efficacy.}
    \label{SI_fig_5}
\end{figure}
\clearpage

\begin{figure}[!htb]
    \centering
    \includegraphics[width=0.9\textwidth]{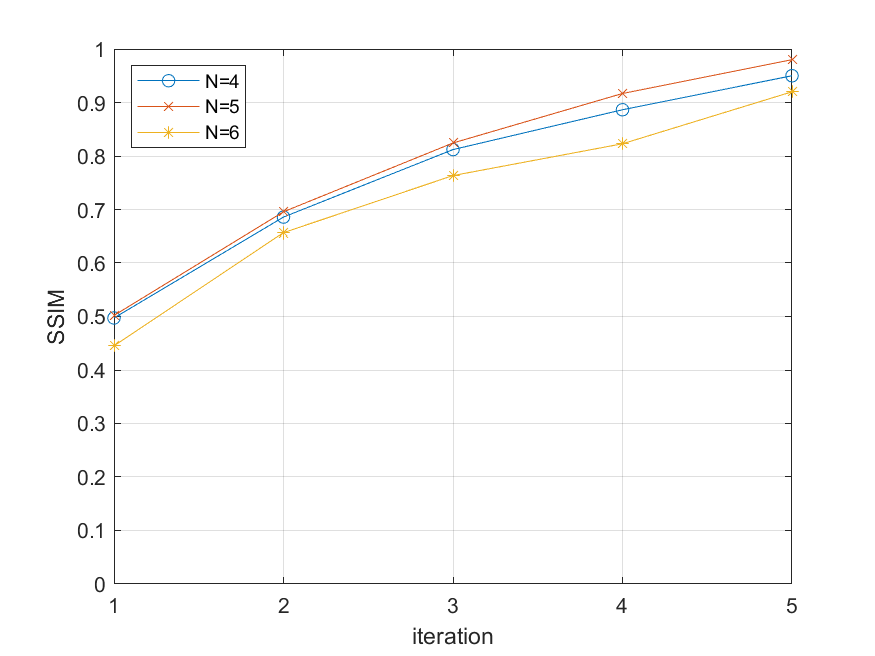}
    \caption{\textbf{Structure similarity index measure for the iterative workflow.} Progression of the similarity metric (SSIM) across five iterations, comparing the refined tilt series to its low-rank component, $\mL$. A higher SSIM value indicates a closer match between the refined series and its low-rank counterpart, suggesting improved quality.}
    \label{SI_fig_6}
\end{figure}
\clearpage

\begin{figure}[!htb]
    \centering
    \includegraphics[width=0.9\textwidth]{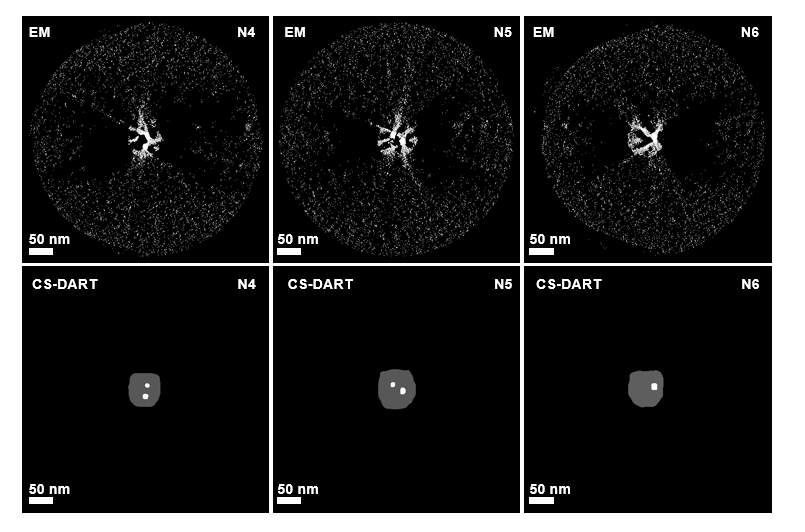}
    \caption{\textbf{Comparison of 3D orthoslices from ML-EM and CS-DART reconstructions of colloidal assemblies.} The top row showcases ML-EM reconstructions, whereas the bottom row presents those from CS-DART. From left to right, the panels correspond to colloidal assemblies containing 4, 5, and 6 particles, respectively.}
    \label{SI_fig_7}
\end{figure}
\clearpage

\begin{figure}[!htb]
    \centering
    \includegraphics[width=0.9\textwidth]{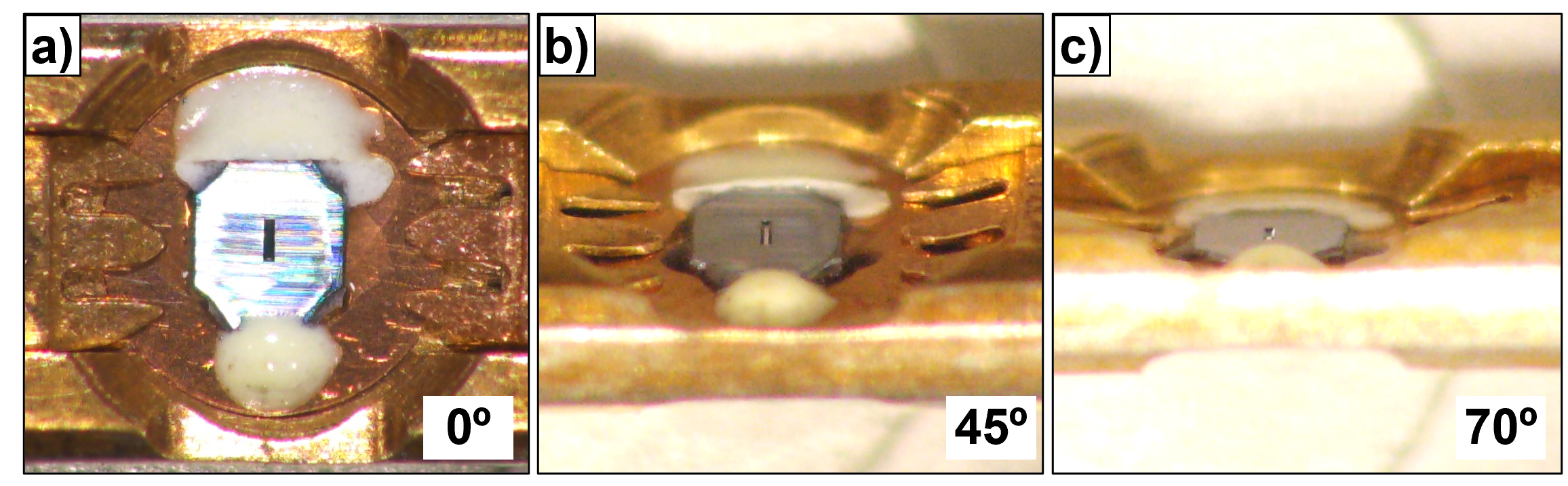}
    \caption{\textbf{Available tilt range of Tomo-chip.} Optical micrographs of a Tomo-chip loaded on a single-tilt tomography holder, with a) $0\degree$, b) $45\degree$, and c) $70\degree$ tilting view, respectively.}
    \label{SI_fig_8}
\end{figure}
\clearpage

\begin{figure}[!htb]
    \centering
    \includegraphics[width=0.9\textwidth]{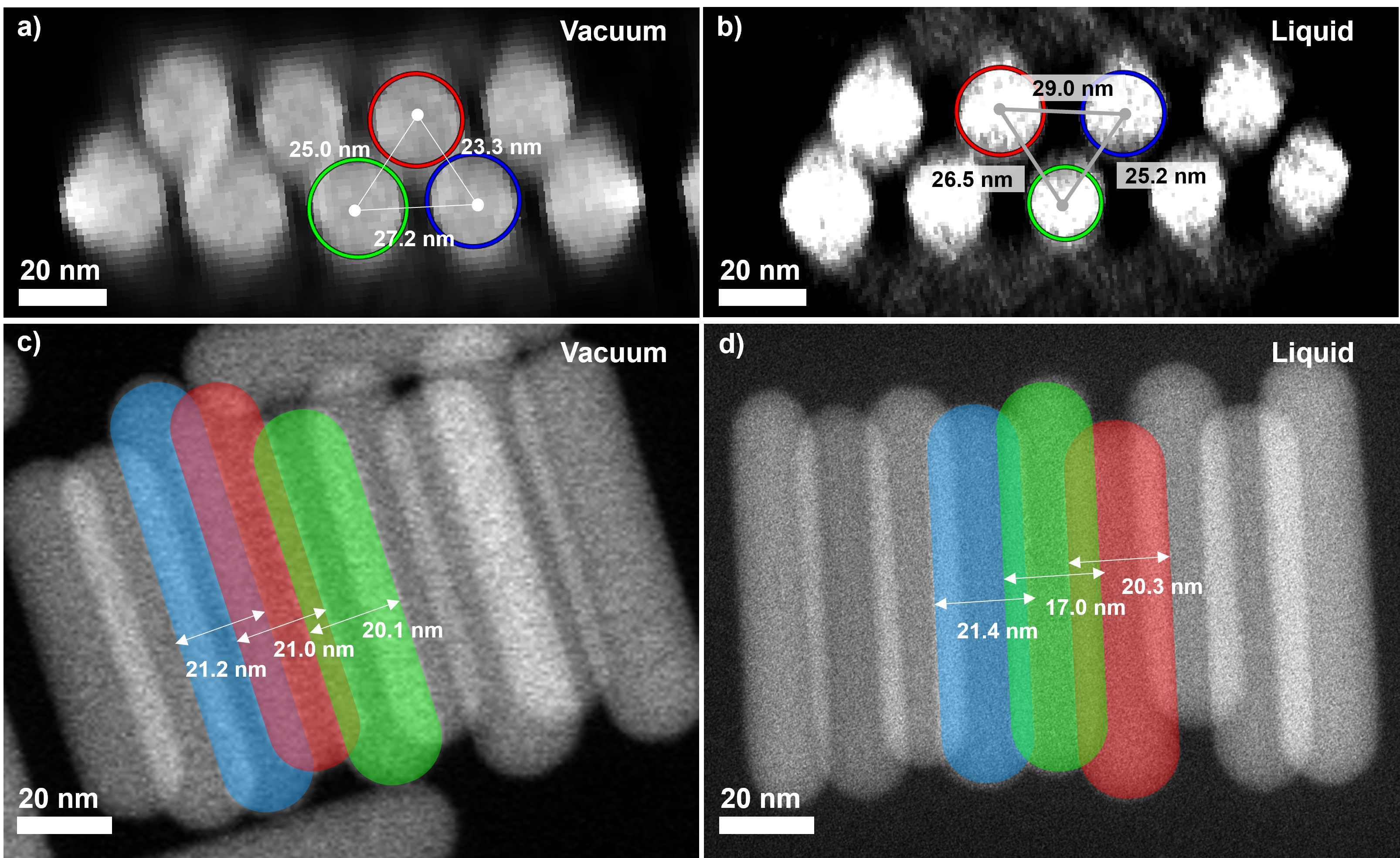}
    \caption{\textbf{Determining the diameters of Au NRs in vacuum and liquid environments.} Projection images captured from the tilt series of the assemblies under investigation in a) vacuum and b) liquid environment, respectively. The individual NRs selected for interparticle measurements were identified within each image, and their diameters were measured independently. This procedure ensured precise interparticle distance measurements. Orthoslices from the reconstructed volume of the assemblies under c) vacuum and d) liquid environment, where the 3 NRs selected for interparticle measurements are highlighted in RGB code, as well as the distances between the center of mass of each NR.}
    \label{SI_fig_9}
\end{figure}
\clearpage

\setcounter{table}{0}

\begin{table}[]
    \centering
    \begin{tabular}{lccc|ccc}
        \toprule
        & \multicolumn{3}{c}{Dry} & \multicolumn{3}{c}{Liquid} \\
        \cmidrule(lr){2-4} \cmidrule(lr){5-7}
        & R & G & B & R & G & B \\
        \midrule
        R & 0 & 4.4 & 2.2 & 0 & 7.9 & 8.2 \\
        G & 4.4 & 0 & 6.6 & 7.9 & 0 & 5.9 \\
        B & 2.2 & 6.6 & 0 & 8.2 & 5.9 & 0 \\
        \bottomrule
    \end{tabular}
    \caption{\textbf{Surface-to-surface distance between Au NRs (R, G, B) under dry and liquid conditions.} All distances are in nm. For coloring of rods (i.e., Red (R), Green(G), Blue(B)), please refer to Supplementary Fig.~\ref{SI_fig_9}.}
    \label{SI_table_1}
\end{table}
\clearpage

\end{document}

%% file: defs.tex
\newcommand{\eg}{{\it e.g.}}              
\newcommand{\ie}{{\it i.e.}}              

\newcommand{\R}{\mathbb{R}}               

\newcommand{\vct}[1]{\boldsymbol{#1}}      

\newcommand{\mtx}[1]{\boldsymbol{#1}}      

\newcommand{\vp}{\vct{p}}
\newcommand{\vq}{\vct{q}}

\newcommand{\vt}{\vct{t}}

\newcommand{\vx}{\vct{x}}
\newcommand{\vy}{\vct{y}}
\newcommand{\vz}{\vct{z}}
\newcommand{\valpha}{\vct{\alpha}}

\newcommand{\vgamma}{\vct{\gamma}}
\newcommand{\vlambda}{\vct{\lambda}}

\newcommand{\vphi}{\vct{\phi}}

\newcommand{\mL}{\mtx{L}}

\newcommand{\mR}{\mtx{R}}
\newcommand{\mS}{\mtx{S}}

\newcommand{\mW}{\mtx{W}}

\newcommand{\mY}{\mtx{Y}}

\newcommand{\mPsi}{\mtx{\Psi}}

